\documentclass[a4paper,12pt]{article}
\usepackage[cp1251]{inputenc}

\usepackage{amsfonts,amssymb,amsthm,mathtools}
\usepackage{amsmath}
\usepackage{icomma}
\usepackage{latexsym}
\usepackage{graphicx}
\usepackage{subfigure}
\usepackage {bm}

\usepackage{geometry} 
\begin{document}
	
	\begin{center}
	\textbf{QUANTUM ELECTRODYNAMICS WITH SELF-CONJUGATED EQUATIONS WITH SPINOR WAVE FUNCTIONS FOR \mbox{FERMION FIELDS}}\\
\end{center}

\begin{center}

		{{V.~P.~Neznamov$^{1,2}$\footnote{vpneznamov@mail.ru, vpneznamov@vniief.ru}, V.~E.~Shemarulin$^{1}$\footnote{VEShemarulin@vniief.ru}}\\

\hfil
	 {\it \mbox{	$^{1}$Russian Federal Nuclear Center--All-Russian Research Institute of Experimental Physics},  Mira pr., 37, Sarov, 607188, Russia \\
		$^{2}$National Research Nuclear University MEPhI, Moscow, 115409, Russia}}
\end{center}
	

\begin{abstract}
\noindent
\footnotesize{Quantum electrodynamics (QED) with self-conjugated equations with spinor wave functions for fermion fields is considered. In the low order of the perturbation theory, matrix elements of some of QED physical processes are calculated. The final results coincide with cross-sections calculated in the standard QED. The self-energy of an electron and amplitudes of processes associated with determination of the anomalous magnetic moment of an electron and Lamb shift are calculated. These results agree with the results in the standard QED. Distinctive feature of the developed theory is the fact that only states with positive energies are present in the intermediate virtual states in the calculations of the electron self-energy, anomalous magnetic moment of an electron and Lamb shift. Besides, in equations, masses of particles and antiparticles have the opposite signs.}\\



\end{abstract}

\section{Introduction}	

In quantum mechanics, motion of half-spin particles is usually described by Dirac equation with first-order derivatives with respect to space-time variables of bispinor wave function  \cite{bib1}.

Motion of fermions can also be described by self-conjugated equations with spinor wave functions  \cite{bib2}.

Under transformation of the Dirac equation with bispinor wave function to self-conjugated equations with spinor wave functions, the energy of the particle (antiparticle) is preserved while probability densities of particle (antiparticle) are different. This leads to new physical consequences. For instance, in case of black holes, nonregular stationary solutions of Dirac equation in external gravitational and electromagnetic Schwarzschild, Reissner-Nordstr\"{o}m, Kerr, Kerr-Newman fields become regular stationary solutions of self-conjugated equations with square-integrable spinor wave functions (see Refs. \cite{bib3}--\cite{bib5}).

The analysis of equations with spinor wave functions in the Coulomb repulsive field has shown existence of the impenetrable potential barrier in the effective potential with the radius proportional to the classical fermion radius and inversely proportional to fermion energy (at  $E\gg mc^{2})$), where $E$  and $m$ are fermion energy and mass  \cite{bib6}. The existence of the impenetrable barrier does not contradict the results of the experiments in probing the internal structure of an electron \cite{bib7} and has no effect on the cross-section of the Coulomb scattering of electrons in the low order of the perturbation theory.

In this paper, we present quantum electrodynamics (QED) with self-conjugated equations with spinor wave functions for fermion fields.

In Sec. 2, we present the self-conjugated equations with spinor wave functions \cite{bib6}. In Sec. 3, we briefly describe the quantum electrodynamics formalism with fermion equations with spinor wave functions. In Sec. 4, we analyze self-energy Feynman diagrams for the QED version under consideration. In Secs. 5, and  6, we calculate matrix elements to determine the anomalous electron moment and Lamb shift by the example of radiative corrections to Coulomb scattering of electrons. In Sec. 7, we discuss the obtained results.

Let us specify these results:
\begin{enumerate}
	\item  In the low order of the perturbation theory, we examine matrix elements 
	of the Coulomb scattering of electrons, scattering of an electron on a 
	proton, Compton effect, and electron-positron pair annihilation. The final 
	results agree with the similar values calculated in the standard QED when 
	using the Dirac equation with the bispinor wave function.
	\item The matrix elements for determining the anomalous magnetic moment of an 
	electron and the Lamb shift agree with the matrix elements determined in the 
	standard QED.
	\item The self--energy of an electron with ultraviolet logarithmic divergence 
	agrees with the value calculated in the standard QED.
	\item Distinctive feature of the developed theory is the fact that only states with positive energies are present in the intermediate virtual states in the calculations of the electron self-energy, anomalous magnetic moment of an electron and Lamb shift. In this case, in the standard QED, the self--energy linearly diverges at the infinite upper limit of integration. Only accounting of the contribution of virtual states with negative energies leads to ultraviolet logarithmic divergence of self-energy (see, for example, Ref. \cite{bib8}).
	\item  In equations, masses of particles and antiparticles have the opposite signs.
	\item Both for electrons and for positrons, in theory there are no transitions between the states with positive and negative energies. In theory, the concept of creation and annihilation of virtual electron-positron pairs is not necessary.
\end{enumerate}

In App. A, operators of interaction $\sim e^{2},e^{3}$ are presented. In App. B, the calculation results for some QED effects are presented. In App. C, the contributions of some diagrams to the Lamb shift of atomic levels are presented. 

In this paper, we use the system of units of $\hslash =c=1$ and the Minkowski space-time signature.
\begin{equation}
	\label{eq1}
	\eta_{\mu \nu } =\mbox{diag}\left[ {1,-1,-1,-1} \right].
\end{equation}

In (\ref{eq1}) $\mu ,\nu =0,1,2,3$.

\section{Self-conjugated equations with spinor wave functions for fermions in the external electromagnetic field}

For fermions, with mass $m$ and charge $e$, moving in the external 
electromagnetic field, the Dirac equation can be written as follows:
\begin{equation}
	\label{eq2}
	\left[ {p^{0}-eA^{0}\left( {{\rm {\bf r}},t} \right)-{\rm {\bm \alpha 
		}}\left( {{\rm {\bf p}}-e{\rm {\bf A}}\left( {{\rm {\bf r}},t} \right)} 
		\right)-\beta m} \right]\psi \left( {{\rm {\bf r}},t} \right)=0.
\end{equation}
Here, $\psi \left( {{\rm {\bf r}},t} \right)$ is a bispinor wave function; 
$A^{0}\left( {{\rm {\bf r}},t} \right),\,\,{\rm {\bf A}}\left( {{\rm {\bf 
			r}},t} \right)$ are potentials of the electromagnetic field; $\alpha 
^{k},\,\beta $ are four-dimensional Dirac matrices, 
$k=1,2,3; \\ p^{0}=i\left( {\partial \mathord{\left/ {\vphantom {\partial 
				{\partial t}}} \right. \kern-\nulldelimiterspace} {\partial t}} 
\right),\,\,{\rm {\bf p}}=-i\vec{{\nabla }}$.

Let
\begin{equation}
	\label{eq3}
	\psi \left( {{\rm {\bf r}},t} \right)=\left( {{\begin{array}{*{20}c}
				{\varphi \left( {{\rm {\bf r}},t} \right)} \hfill \\
				{\chi \left( {{\rm {\bf r}},t} \right)} \hfill \\
	\end{array} }} \right),
\end{equation}
where $\varphi \left( {{\rm {\bf r}},t} \right),\,\,\chi \left( {{\rm {\bf 
			r}},t} \right)$ are spinor wave functions. Then, the following expressions 
follow from the equation (\ref{eq2})
\begin{equation}
	\label{eq4}
	\begin{array}{l}
		\left( {p^{0}-eA^{0}-m} \right)\varphi ={\rm {\bm \sigma }}\left( {{\rm 
				{\bf p}}-e{\rm {\bf A}}} \right)\chi , \\ [10pt]
		\left( {p^{0}-eA^{0}+m} \right)\chi ={\rm {\bm \sigma }}\left( {{\rm {\bf 
					p}}-e{\rm {\bf A}}} \right)\varphi , \\ 
	\end{array}
\end{equation}
\begin{equation}
	\label{eq5}
	\begin{array}{l}
		\chi =\dfrac{1}{p^{0}-eA^{0}+m}{\rm {\bm \sigma }}\left( {{\rm {\bf 
					p}}-e{\rm {\bf A}}} \right)\varphi , \\ [8pt]
		\varphi =\dfrac{1}{p^{0}-eA^{0}-m}{\rm {\bm \sigma }}\left( {{\rm {\bf 
					p}}-e{\rm {\bf A}}} \right)\chi . \\ 
	\end{array}
\end{equation}
Here, $\sigma^{i}$ are Pauli matrices. From the equalities (\ref{eq4}) and (\ref{eq5}), 
we can obtain the equations of
\begin{equation}
	\label{eq6}
	\left[ {p^{0}-eA^{0}-m-{\rm {\bm \sigma }}\left( {{\rm {\bf p}}-e{\rm {\bf 
					A}}} \right)\frac{1}{p^{0}-eA^{0}+m}{\rm {\bm \sigma }}\left( {{\rm {\bf 
					p}}-e{\rm {\bf A}}} \right)} \right]\varphi =0,
\end{equation}
\begin{equation}
	\label{eq7}
	\left[ {p^{0}-eA^{0}+m-{\rm {\bm \sigma }}\left( {{\rm {\bf p}}-e{\rm {\bf 
					A}}} \right)\frac{1}{p^{0}-eA^{0}-m}{\rm {\bm \sigma }}\left( {{\rm {\bf 
					p}}-e{\rm {\bf A}}} \right)} \right]\chi =0.
\end{equation}
Equation (\ref{eq6}) differs from equation (\ref{eq7}) by the replacement of $p^{0}\to 
-p^{0}, \\ {\rm {\bf p}}\to -{\rm {\bf p}},e\to -e$ or by the replacement of 
$+m\to -m$.

If we multiply Eq. (\ref{eq6}) on the left by operator $\left( 
{p^{0}-eA^{0}+m} \right)$, and Eq. (\ref{eq7}) -- by operator $\left( 
{p^{0}-eA^{0}-m} \right)$, we will obtain that
\begin{equation}
	\label{eq8}
		\begin{array}{l}
	\left[ {\left( {p^{0}-eA^{0}} \right)^{2}-m^{2}-\left( {p^{0}-eA^{0}+m} 
		\right){\rm {\bm \sigma }}\left( {{\rm {\bf p}}-e{\rm {\bf A}}} 
		\right)} \right. \\ [8pt]
	\left. { \times \dfrac{1}{p^{0}-eA^{0}+m}{\rm {\bm \sigma }}\left( {{\rm {\bf 
					p}}-e{\rm {\bf A}}} \right)} \right]\varphi =0,
		\end{array}
\end{equation}
\begin{equation}
	\label{eq9}
		\begin{array}{l}
	\left[ {\left( {p^{0}-eA^{0}} \right)^{2}-m^{2}-\left( {p^{0}-eA^{0}-m} 
		\right){\rm {\bm \sigma }}\left( {{\rm {\bf p}}-e{\rm {\bf A}}} 
		\right) } \right. \\ [8pt]
	\left. { \times \dfrac{1}{p^{0}-eA^{0}-m}{\rm {\bm \sigma }}\left( {{\rm {\bf 
					p}}-e{\rm {\bf A}}} \right)} \right]\chi =0.
		\end{array}
\end{equation}
Equations (\ref{eq8}) and (\ref{eq9}) must be reduced to the self-conjugated form 
\cite{bib6}--\cite{bib10}. It is worth to note that the similarity transformation operator is written in the closed 
form of
\begin{equation}
	\label{eq10}
		\Phi =g_{\varphi } \varphi , \,\,\,\,\,{\rm X}=g_{\chi } \chi ,  
\end{equation}
where
\begin{equation}
	\label{eq11}
		g_{\varphi } =\left( {p^{0}-eA^{0}+m} \right)^{{-1/2}}, \,\,\,\,\,\,
		g_{\chi } =\left( {p^{0}-eA^{0}-m} \right)^{{-1/2}}.  
\end{equation}
After the transformations, Eqs. (\ref{eq8}) and (\ref{eq9}) assume the self-conjugated 
form of
\begin{equation}
	\label{eq12}
	\begin{array}{l}
\left[ {\left( {p^{0}-eA^{0}} \right)^{2}-m^{2}-\left( {p^{0}-eA^{0}+m} 
	\right)^{1/2}{\rm {\bm \sigma }}\left( {{\rm {\bf p}}-e{\rm 
			{\bf A}}} \right) } \right. \\ [5pt]
\left. {\times \dfrac{1}{p^{0}-eA^{0}+m}{\rm {\bm \sigma }}\left( {{\rm {\bf 
				p}}-e{\rm {\bf A}}} \right)\left( {p^{0}-eA^{0}+m} \right)^{1/2}} 
\right]\Phi=0, 
		\end{array}
\end{equation}
\begin{equation}
	\label{eq13}
\begin{array}{l}
	\left[ {\left( {p^{0}-eA^{0}} \right)^{2}-m^{2}-\left( {p^{0}-eA^{0}-m} 
		\right)^{1/2}{\rm {\bm \sigma }}\left( {{\rm {\bf p}}-e{\rm 
				{\bf A}}} \right) } \right. \\ [5pt]
	\left. {\times \dfrac{1}{p^{0}-eA^{0}-m}{\rm {\bm \sigma }}\left( {{\rm {\bf 
					p}}-e{\rm {\bf A}}} \right)\left( {p^{0}-eA^{0}-m} \right)^{1/2}} 
	\right] X=0, 
\end{array}
\end{equation}

As before, Eqs. (\ref{eq12}) and (\ref{eq13}) transform to each other at the replacement 
of $p^{0}\to -p^{0}$, ${\rm {\bf p}}\to -{\rm {\bf p}},\,\,e\to -e$ or at 
the replacement of $+m\to -m$.

\subsection{C, P, T- symmetries}

Equations (\ref{eq12}) and (\ref{eq13}) is $P$-invariant at the space reflection ${\rm {\bf 
		{x}'}}=-{\rm {\bf x}}$.

Operations of C- and T-conjugation are reduced to the operation of the 
complex conjugation in both cases.
\[
\begin{array}{l}
	\Phi^{C}=\Phi^ \ast ,\,\,\,\,\Phi^{T}=\Phi ^\ast , \\ [10pt]
	{\rm X}^{C}={\rm X^\ast} ,\,\,\,\,{\rm X}^{T}={\rm X^\ast} . \\ 
\end{array}
\]
An equation for $\Phi^{C}$ and ${\rm X}^{C}$ has the form of
\begin{equation}
	\label{eq14}
	\begin{array}{l}
		\left[ {\left( {p^{0}+eA^{0}} \right)^{2}-m^{2}-\left( {p^{0}+eA^{0}-m} 
			\right)^{1/2}{\rm {\bm \sigma }}\left( {{\rm {\bf p}}+e{\rm 
					{\bf A}}} \right) } \right. \\ [5pt]
		\left. {\times \dfrac{1}{p^{0}+eA^{0}-m}{\rm {\bm \sigma }}\left( {{\rm {\bf 
						p}}+e{\rm {\bf A}}} \right)\left( {p^{0}+eA^{0}-m} \right)^{1/2}} 
		\right]\Phi^{C}=0, \\ 
	\end{array}
\end{equation}
\begin{equation}
	\label{eq15}
	\begin{array}{l}
		\left[ {\left( {p^{0}+eA^{0}} \right)^{2}-m^{2}-\left( {p^{0}+eA^{0}+m} 
			\right)^{1/2}{\rm {\bm \sigma }}\left( {{\rm {\bf p}}+e{\rm 
					{\bf A}}} \right) } \right. \\ [5pt]
		\left. {\times \dfrac{1}{p^{0}+eA^{0}+m}{\rm {\bf \sigma }}\left( {{\rm {\bf 
						p}}+e{\rm {\bf A}}} \right)\left( {p^{0}+eA^{0}+m} \right)^{1/2}} 
		\right]{\rm X}^{C}=0. \\ 
	\end{array}
\end{equation}
It is easy to see that theory is CPT-invariant. Eq. (\ref{eq14}) for $\Phi 
^{C}$ differs from (\ref{eq12}) by the signs before electric charge and mass. If Eq. (\ref{eq12}) describes the electrons with charge $-e$ and mass $+m$, then Eq. (\ref{eq14}) for $\Phi^{C}$ describes the positrons with charge $+e$ 
and mass $-m$.

\subsection{\mbox{Choice of the equations for the quantum electrodynamics}}

Each of Eqs. (\ref{eq12})--(\ref{eq15}) contains solutions with 
positive and negative energies. In the following, in calculations, we will use only 
solutions with positive energy. Solutions with negative energy are used only 
for the mathematical completeness.

We will use solutions of Eqs. (\ref{eq12}) and (\ref{eq14}) with positive energies 
for electron and positron. Solutions of Eqs. (\ref{eq13}) and (\ref{eq15}) 
correspond to solutions for electrons with mass $-m$ and solutions for 
positrons with mass $+m$. These solutions do not contain additional physical information compared to the solutions of Eqs. (\ref{eq12}) and (\ref{eq14}), so, below, we will not consider them.

Then, in QED, spinor $\Phi \left( {x,s} \right)$ will 
represent the operator of electron field with positive energy and mass $+m$, 
spinor $\Phi^{C}\left( {x,s} \right)$ will represent the operator of 
positron field with positive energy and mass $-m$.

In case of absence of electromagnetic field $\left( {A^{\mu }=0} \right)$ , 
Eqs. (\ref{eq12}) and (\ref{eq14}) represent free Klein-Gordon equations with spinor wave 
functions
\begin{equation}
	\label{eq16}
	\left( {p_{0}^{2} -{\rm {\bf p}}^{2}-m^{2}} \right)F_{0}^{\pm } \left( {x,s} 
	\right)=0.
\end{equation}
Here, $F_{0}^{+} \left( {x,s} \right)=\Phi_{0} \left( {x,s} \right)=\Phi 
_{0}^{C} \left( {x,s} \right)$.

Solutions (\ref{eq16}) in the form of plane waves normalized in the continuous 
spectrum have the form of
\begin{equation}
	\label{eq17}
	F_{0}^{\pm } \left( {x,p,s} \right)=\frac{1}{\sqrt {\left( {2\pi } 
			\right)^{3}2E_{p} } }e^{\mp ipx}U_{s} ,
\end{equation}
where $E_{p} =p^{0}>0$ and $p^{2}=m^{2}$, $U_{s} $ are normalized 
two-component Pauli spin functions.

The conditions of orthonormalization are:
\begin{equation}
	\label{eq18}
	\begin{array}{l}
		\int {d{\rm {\bf x}}\bar{F}_{0}^{\pm } \left( {x,{p}',{s}'} 
			\right)i\mathord{\buildrel{\lower3pt\hbox{$\scriptscriptstyle\leftrightarrow$}}\over 
				{\partial }}_{0} } F_{0}^{\pm } \left( {x,p,s} \right)=\pm \,\delta \left( 
		{{\rm {\bf p}}-{\rm {\bf {p}'}}} \right)\delta_{s{s}'} , \\ [10pt]
		\int {d{\rm {\bf x}}\bar{{F}}_{0}^{\pm } \left( {x,{p}',{s}'} 
			\right)i\mathord{\buildrel{\lower3pt\hbox{$\scriptscriptstyle\leftrightarrow$}}\over 
				{\partial }}_{0} } F_{0}^{\mp } \left( {x,p,s} \right)=0, \\ 
	\end{array}
\end{equation}
where denomination of 
$a \mathord{\buildrel{\lower3pt\hbox{$\scriptscriptstyle\leftrightarrow$}}\over 
	{\partial }}_{0} b\equiv a\frac{\partial b}{\partial t}-\frac{\partial 
	a}{\partial t}b$. In (\ref{eq18}), the line over the function means Hermitian 
conjugation.

\section{Formalism of quantum electrodynamics with fermion equations with spinor wave functions}

\subsection{Propagator for Klein-Gordon equation}

The Feynman propagator is found from the solution of the equation of
\begin{equation}
	\label{eq19}
	\left( {\Box_{{x}'} +m^{2}} \right)\Delta_{F} \left( {{x}'-x} \right)=-\delta 
	^{4}\left( {{x}'-x} \right),
\end{equation}
where $\Box_{{x}'} $ is D'Alembertian.

With transition to momentum representation, we obtain (see, for instance, \mbox{Ref. \cite{bib9})}
\begin{equation}
	\label{eq20}
	\Delta_{F} \left( {{x}'-x} \right)=\int {d^{4}p\frac{e^{-ip\left( {{x}'-x} 
				\right)}}{\left( {2\pi } \right)^{4}}\frac{1}{p^{2}-m^{2}+i\varepsilon }} .
\end{equation}
In our theory, in propagator (\ref{eq20}), we use only pole with positive energy.

\subsection{Interaction operators}

Let us rewrite Eqs. (\ref{eq12}) and (\ref{eq14}) as
\begin{equation}
	\label{eq21}
	\left( {\left( {p^{0}} \right)^{2}-{\rm {\bf p}}^{2}-m^{2}} \right)\Phi 
	\left( {x,s} \right)=V\left( {p^{\mu },-eA^{\mu },+m} \right)\Phi \left( 
	{x,s} \right),
\end{equation}
\begin{equation}
	\label{eq22}
	\left( {\left( {p^{0}} \right)^{2}-{\rm {\bf p}}^{2}-m^{2}} \right)\Phi 
	^{C}\left( {x,s} \right)=V\left( {p^{\mu },eA^{\mu },-m} \right)\Phi 
	^{C}\left( {x,s} \right),
\end{equation}
where
\begin{eqnarray}
	\label{eq23}
	\begin{array}{l}
		V\left( {p^{\mu },\mp eA^{\mu },\pm m} \right)=\pm e\left( 
		{p^{0}A^{0}+A^{0}p^{0}} \right)-e^{2}A_{0}^{2} -{\rm {\bf p}}^{2} \\ [10pt]
		+ \left( {p^{0}\mp eA^{0}\pm m} \right)^{1/2}{\rm {\bm \sigma }}\left( {{\rm {\bf p}}\mp 
			e{\rm {\bf A}}} \right) \\ [8pt]
	\times \dfrac{1}{p^{0}\mp eA^{0}\pm m}{\rm {\bm \sigma }}\left( {{\rm {\bf 
					p}}\mp e{\rm {\bf A}}} \right)\left( {p^{0}\mp eA^{0}\pm m} \right)^{1/2}. \\ 
	\end{array}
\end{eqnarray}
Here, the upper sign corresponds to Eq. (\ref{eq12}), the lower sign 
corresponds to Eq. (\ref{eq14}).

As against the Dirac QED representation, our representation leads to the 
infinite set of interaction vertices depending on the order of the 
perturbation theory.

In this paper, we will consider QED processes $\sim 
\,e,\,e^{2}$ and $e^{3}$. Let us expand expression (\ref{eq23}) to the power of 
$\sim \,e^{3}$.

For convenience, let us pass over to the representation where momentum 
variables are diagonal. In this representation, the matrix element is 
\begin{equation}
	\label{eq24}
	\left\langle {{\rm {\bf {p}'}}\left| {e^{i{\rm {\bf kx}}}} \right|{\rm {\bf 
				{{p}''}}}} \right\rangle =\delta \left( {{\rm {\bf {p}'}}-{\rm {\bf 
				{{p}''}}}-{\rm {\bf k}}} \right).
\end{equation}
Let us represent the field of $A^{\mu }\left( {{\rm {\bf x}},t} \right)$ 
with the Fourier integral. For the case of QED, in 
denotations of Ref. \cite{bib11}, we have
\begin{equation}
	\label{eq25}
	A^{\mu }\left( {{\rm {\bf x}},t} \right)=\sum\limits_{\nu=\pm 1} {\int 
		{A_{\left( \nu \right)k}^{\mu } } \left( t \right)e^{-i{\rm {\bf kx}}}d{\rm 
			{\bf k}}} ,
\end{equation}
where
\begin{equation}
	\label{eq26}
	A_{\left( \nu \right)k}^{\mu } \left( t \right)=\left\{ {\begin{array}{l}
			A_{k}^{\mu } e^{ik_{0} t},\,\,\,\,\,\,\,\,\,\,\,\,\nu=1, \\ [5pt]
			\bar{{A}}_{\left( {-k} \right)}^{\mu } e^{-ik_{0} t},\,\,\nu=-1. \\ 
	\end{array}} \right.
\end{equation}
In the representation of (\ref{eq24})
\begin{equation}
	\label{eq27}
	\left\langle {{\rm {\bf {p}'}}\left| {A^{\mu }} \right|{\rm {\bf {{p}''}}}} 
	\right\rangle =\sum\limits_{\nu=\pm 1} {A_{\left( \nu \right)\left( 
			{{{p}''}-{p}'} \right)}^{\mu } \left( t \right);\,\,\,\left| {{\rm {\bf k}}} 
		\right|=\left| {{\rm {\bf {{p}''}}}-{\rm {\bf {p}'}}} \right|} ,
\end{equation}
\begin{equation}
	\label{eq28}
	\begin{array}{l}
				\left\langle {{\rm {\bf {p}'}}\left| {A^{\mu }A^{\lambda }} \right|{\rm 
				{\bf {{p}''}}}} \right\rangle  = \int {d{\rm {\bf {{{p}'''}}}}} \left\langle 
		{{\rm {\bf {p}'}}\left| {A^{\mu }} \right|{\rm {\bf {{{p}'''}}}}} 
		\right\rangle \left\langle {{\rm {\bf {{{p}'''}}}}\left| {A^{\lambda }} 
			\right|{\rm {\bf {{p}''}}}} \right\rangle \\ [12pt]
		=\sum\limits_{\nu,{\nu}' = \pm 1} {d{\rm {\bf {{{p}'''}}}}A_{\left( \nu 
				\right)\left( {{{{p}'''}}-{p}'} \right)}^{\mu } \left( t \right)A_{\left( 
				{{\nu}'} \right)\left( {{{p}''}-{{{p}'''}}} \right)}^{\lambda } \left( t 
			\right)}; \\ [12pt]
			{\left| {{\rm {\bf k}}} \right|=\left| {{\rm {\bf 
					{{{p}'''}}}}-{\rm {\bf {p}'}}} \right|} ,\,\,\left| {{\rm {\bf {k}'}}} 
		\right|=\left| {{\rm {\bf {{p}''}}}-{\rm {\bf {{{p}'''}}}}} \right|, \\ 
	\end{array}
\end{equation}
\begin{equation}
	\label{eq29}
	\begin{array}{l}
		\left\langle {{\rm {\bf {p}'}}\left| {A^{\mu }A^{\lambda }A^{\delta }} 
			\right|{\rm {\bf {{p}'}'}}} \right\rangle =\sum\limits_{\nu,{\nu}',{{\nu}''}=\pm 
			1} {\int {d{\rm {\bf {{{p}'''}}}}d{\rm {\bf p}}^{IV}} A_{\left( \nu 
				\right)\left( {{{{p}'''}}-{p}'} \right)}^{\mu } \left( t \right)A_{\left( 
				{{\nu}'} \right)\left( {p^{IV}-{{{p}'''}}} \right)}^{\lambda } \left( t 
			\right)\,} \\ [15pt]
		\times A_{\left( {{{\nu}''}} \right)\left( {{{p}''}-p^{IV}} \right)}^{\delta 
		} \left( t \right);\\ [15pt]
	\left| {{\rm {\bf k}}} \right|=\left| {{\rm {\bf 
					{{{p}'''}}}}-{\rm {\bf {p}'}}} \right|,\,\,\,\left| {{\rm {\bf {k}'}}} 
		\right|=\left| {{\rm {\bf p}}^{IV}-{\rm {\bf {{{p}'''}}}}} 
		\right|,\,\,\left| {{\rm {\bf {{k}''}}}} \right|=\left| {{\rm {\bf 
					{{p}''}}}-{\rm {\bf p}}^{IV}} \right|. \\ 
	\end{array}
\end{equation}
Let us denote $X=p^{0}\pm m$, where operator $p^{0}=i\frac{\partial 
}{\partial t}$ as before. Then, in (\ref{eq23}),
\begin{equation}
	\label{eq30}
	\begin{array}{l}
		\dfrac{1}{p^{0}\pm m\mp eA^{0}}=\dfrac{1}{X}+\dfrac{1}{X}\left( {\pm e} 
		\right)A^{0}\dfrac{1}{X}+\dfrac{1}{X}\left( {\pm e} 
		\right)A^{0}\dfrac{1}{X}\left( {\pm e} \right)A^{0}\dfrac{1}{X} \\ [10pt]
		+\dfrac{1}{X}\left( {\pm e} \right)A^{0}\dfrac{1}{X}\left( {\pm e} 
		\right)A^{0}\dfrac{1}{X}\left( {\pm e} \right)A^{0}\dfrac{1}{X}+... \\ 
	\end{array}
\end{equation}
At expansion of $\left( {p^{0}\mp eA^{0}\pm m} \right)^{1/2}$ we will use 
representation (\ref{eq24})
\begin{equation}
	\label{eq31}
	\left( {X\mp eA^{0}} \right)^{1/2}=X^{1 \mathord{\left/ {\vphantom {1 2}} \right. 
			\kern-\nulldelimiterspace} 2}+A+B+C.
\end{equation}
Here, the operators are $A\sim e,\,\,B\sim e^{2},\,\,C\sim e^{3}$
\[
X\mp eA^{0}=\left( {X^{1/2}+A+B+C} \right)^{2}.
\]
It implies that

\begin{enumerate}
	\item $AX^{1/2}+X^{1/ 2}A=\mp eA^{0},$
	\begin{equation}
		\label{eq32}
		\left\langle {{\rm {\bf {p}'}}\left| A \right|{\rm {\bf {{p}''}}}} 
		\right\rangle =\left\langle {{\rm {\bf {p}'}}\left| {\frac{\left( {\mp e} 
					\right)A^{0}}{I+II}} \right|{\rm {\bf {{p}''}}}} \right\rangle,
	\end{equation}
where $I=\left( {{X}'} \right)\,^{1/ 2},\,\,II=\left( {{{X}''}\,} \right)^{1/ 2}.$ In 
	(\ref{eq32}) and below we use $\left\langle {{\rm {\bf p}}'\left| {AX^{1 
			/2}} 
		\right|{\rm {\bf p}}''} \right\rangle =\left\langle {{\rm {\bf p}}'\left| A 
		\right|{\rm {\bf p}}''} \right\rangle \left( {X''} \right)^{1 
		/ 2}$, 
	$\left\langle {{\rm {\bf p}}'\left| {X^{1/2}A} \right|{\rm {\bf p}}''} 
	\right\rangle =\left( {X'} \right)^{1/ 2}\left\langle {{\rm {\bf p}}'\left| A 
		\right|{\rm {\bf p}}''} \right\rangle $, etc.
	\item $BX^{1/2}+X^{1/ 2}B+A^{2}=0,$
	\begin{equation}
		\label{eq33}
		\left\langle {{\rm {\bf {p}'}}\left| B \right|{\rm {\bf {{p}''}}}} 
		\right\rangle =-\int {d{\rm {\bf {{{p}'''}}}}} \frac{\left\langle {{\rm {\bf 
						{p}'}}\left| {\left( {\pm e} \right)A^{0}} \right|{\rm {\bf {{{p}'''}}}}} 
			\right\rangle \left\langle {{\rm {\bf {{{p}'''}}}}\left| {\left( {\pm e} 
					\right)A^{0}} \right|{\rm {\bf {{p}''}}}} \right\rangle }{\left( {I+II} 
			\right)\left( {I+III} \right)\left( {III+II} \right)}.
	\end{equation}
	\item $CX^{1/ 2}+X^{1/ 2}C+BA+AB=0,$
	\begin{equation}
		\label{eq34}
		\begin{array}{l}
		\left\langle {{\rm {\bf {p}'}}\left| C \right|{\rm {\bf {{p}''}}}} 
		\right\rangle =-\int {d{\rm {\bf {{{p}'''}}}}} d{\rm {\bf 
				p}}^{IV} \\ [10pt]
			\times \dfrac{\left( {I+III+IV+II} \right)\left\langle {{\rm {\bf 
						{p}'}}\left| {\left( {\pm e} \right)A^{0}} \right|{\rm {\bf {{{p}'''}}}}} 
			\right\rangle \left\langle {{\rm {\bf {{{p}'''}}}}\left| {\left( {\pm e} 
					\right)A^{0}} \right|{\rm {\bf p}}^{IV}} \right\rangle \left\langle {{\rm 
					{\bf p}}^{IV}\left| {\left( {\pm e} \right)A^{0}} \right|{\rm {\bf 
						{{p}''}}}} \right\rangle }{\left( {I+II} \right)\left( {I+III} \right)\left( 
			{I+IV} \right)\left( {III+II} \right)\left( {IV+II} \right)}.
			\end{array}
	\end{equation}

	In (\ref{eq33}) and (\ref{eq34}) $III=\left( {{{{X}'''}}} \right)\,^{1/ 2},\,\,\,IV=\left( 
	{X^{IV}} \right)\,^{1/ 2}$.
	
	Taking into account (\ref{eq30}) - (\ref{eq34}), sought expansion (\ref{eq23}) can be presented as: 
	
	$V=V_{1} +V_{2} +V_{3} ,$ where $V_{1} \sim \pm e,\,\,V_{2} \sim 
	e^{2},\,\,V_{3} \sim \pm e^{3}$.
	
	In the momentum representation,
	\begin{equation}
		\label{eq35}
		\begin{array}{l}
			\left\langle {{\rm {\bf {p}'}}\left| {V_{1} } \right|{\rm {\bf {{p}''}}}} 
			\right\rangle =\pm e\left[ {\left( {\left( {{p}'} \right)^{0\,\,}+\left( 
					{{{p}''}} \right)^{0\,\,}-\dfrac{\left( {{\rm {\bf {p}'}}} 
						\right)^{2}}{I\left( {I+II} \right)}+\dfrac{{\rm {\bm \sigma \bf {p}' \bm\sigma 
								{{p}''}}}}{I\cdot II}-\dfrac{\left( {{\rm {\bf {{p}''}}}} 
						\right)^{2}}{II\left( {I+II} \right)}} \right) } \right. \\
					\left. { \times  \left\langle {{\rm {\bf 
						{p}'}}\left| {A^{0}} \right|{\rm {\bf {{p}''}}}} \right\rangle +
			\left( {-\dfrac{II}{I}{\rm {\bm \sigma \bf {p}'}} \sigma 
					^{i}-\dfrac{I}{II}\sigma^{i}{\rm {\bm \sigma \bf  {{p}''}}}} \right)\left\langle 
				{{\rm {\bf {p}'}}\left| {A^{i}} \right|{\rm {\bf {{p}''}}}} \right\rangle } 
			\right]. \\ 
		\end{array}
	\end{equation}
	The expressions for $V_{2} ,\,V_{3} $, having a more cumbersome form, are 
	presented in App. A. For convenience, in (\ref{eq35}) and in the formulas of 
	App. A, the factors independent of $A^{\mu }_{\, }$ are taken outside 
	the matrix element.
\end{enumerate}

\subsection{Feynman rules}

Feynman rules can be determined by means of propagator method \cite{bib9} and comparison with Feynman rules for scalar electrodynamics of spinless charged particles (see, for example, Ref. \cite{bib9}). Reference \cite{bib17} contains the more strict definition of Feynman rules for electrodynamics with spinors in the fermion equations and with Dirac matrixes in spinor 
representation.

As against the scalar electrodynamics, in our case, there exists the 
infinite set of interaction vertices with photon depending on the order of 
the perturbation theory (see (\ref{eq23})): factor $\left( {-iV_{1\mu } } \right)$ 
corresponds to the vertex of interaction with one photon, factor $\left( 
{-iV_{2\mu \nu } } \right)$ corresponds to the vertex of interaction with 
two photons, etc. For convenience, values $V_{1\mu } ,\,\,V_{2\mu \nu } 
,\,...$ stand for the appropriate terms of the interaction operator (\ref{eq23}) 
without electromagnetic potentials $A^{\mu },A^{\mu }A^{\nu },\,\,...$

Each external fermion line corresponds function $F_{0}^{+} \left( {x,p,s} 
\right)$ (see (\ref{eq17})).

The rest of the Feynman rules are the same as in scalar electrodynamics of 
charged bosons.

\subsection{Calculations of QED processes}

Taking into account the formulated Feynman rules, we considered some of the 
QED processes in the first, second and third orders of the perturbation 
theory. We calculated the matrix elements of Coulomb scattering of 
electrons, the M{\o}ller scattering, Compton effect, and annihilation of 
electron-positron pair as well as the matrix elements to determine 
self-energy of an electron, anomalous magnetic moment of electrons, Lamb 
shift of atomic energy levels.

Figures 1--4 present Feynman diagrams of some of the processes under 
consideration. Some of the computational details are given in App. B.

The final computational results of the QED amplitudes whose diagrams are 
given in Figs. 1-3 agree with similar values calculated in Dirac 
representation.
\begin{figure}[!ht]
	\begin{center}
			\includegraphics[width=0.3\linewidth]{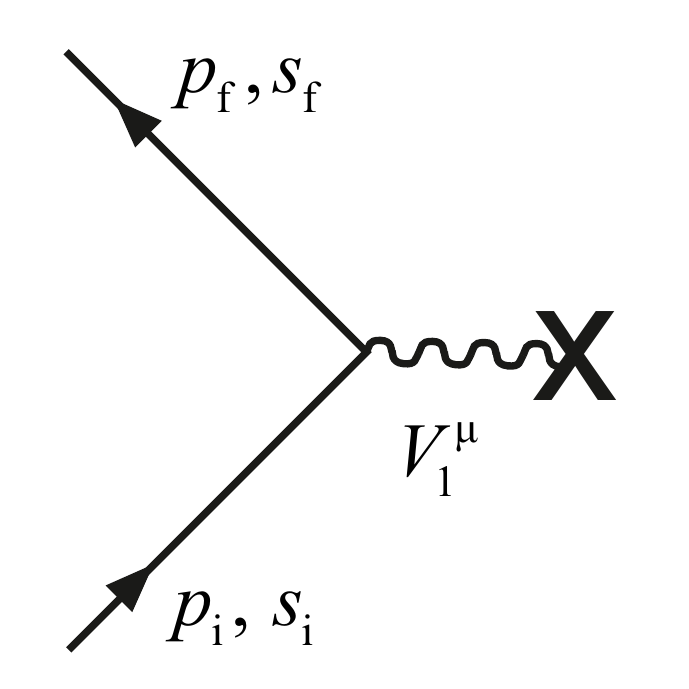}
			\caption{Electron scattering in the Coulomb field} 
			\label{Fig.1} 
		\end{center}
		\end{figure}
\begin{figure}[!ht]
	\begin{center}
			\includegraphics[width=0.5\linewidth]{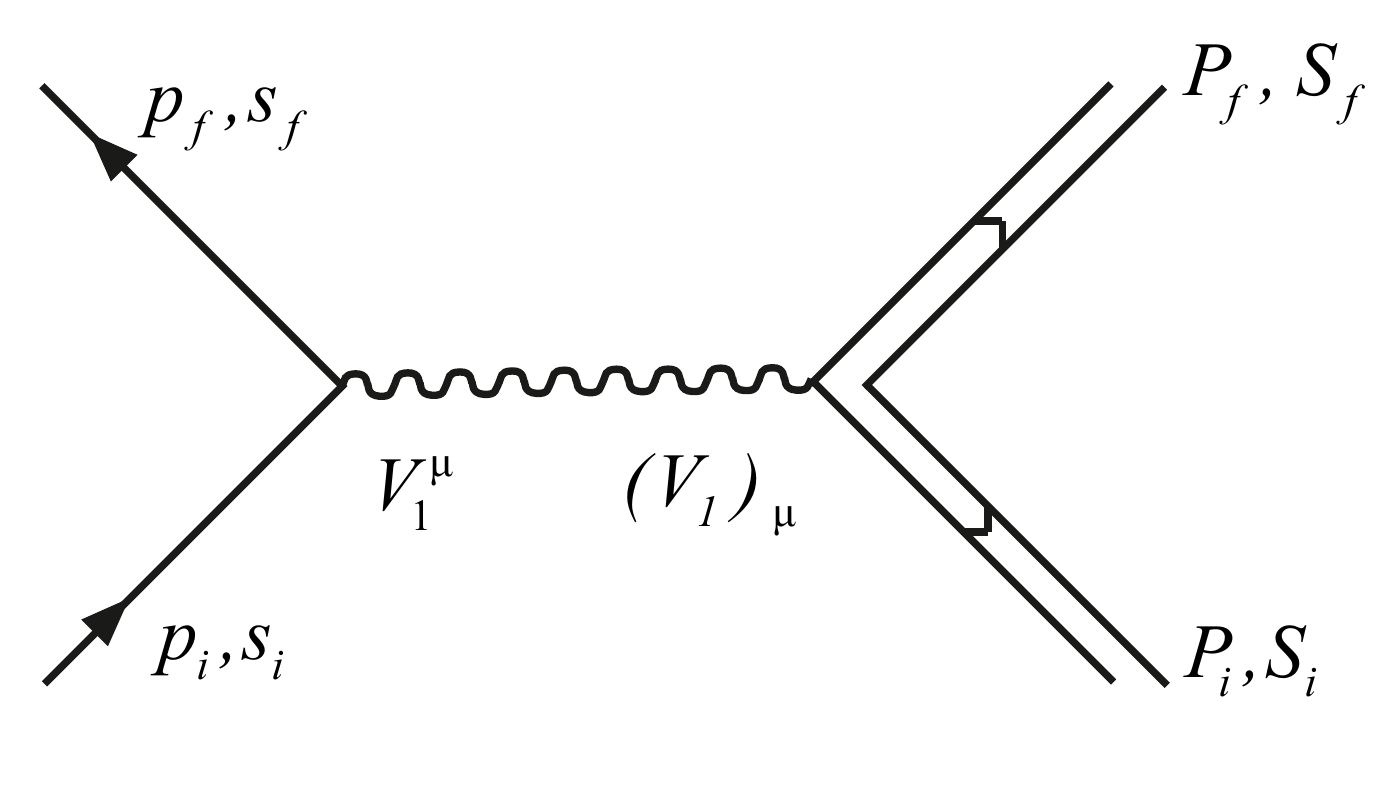}
			\caption{Scattering of an electron on a Dirac proton (M{\o}ller scattering)}
			\label{Fig.2}
			\end{center}
\end{figure}

\begin{figure}[h!]
	\begin{minipage}[h]{0.45\linewidth}
		\center{\includegraphics[width=1\linewidth]{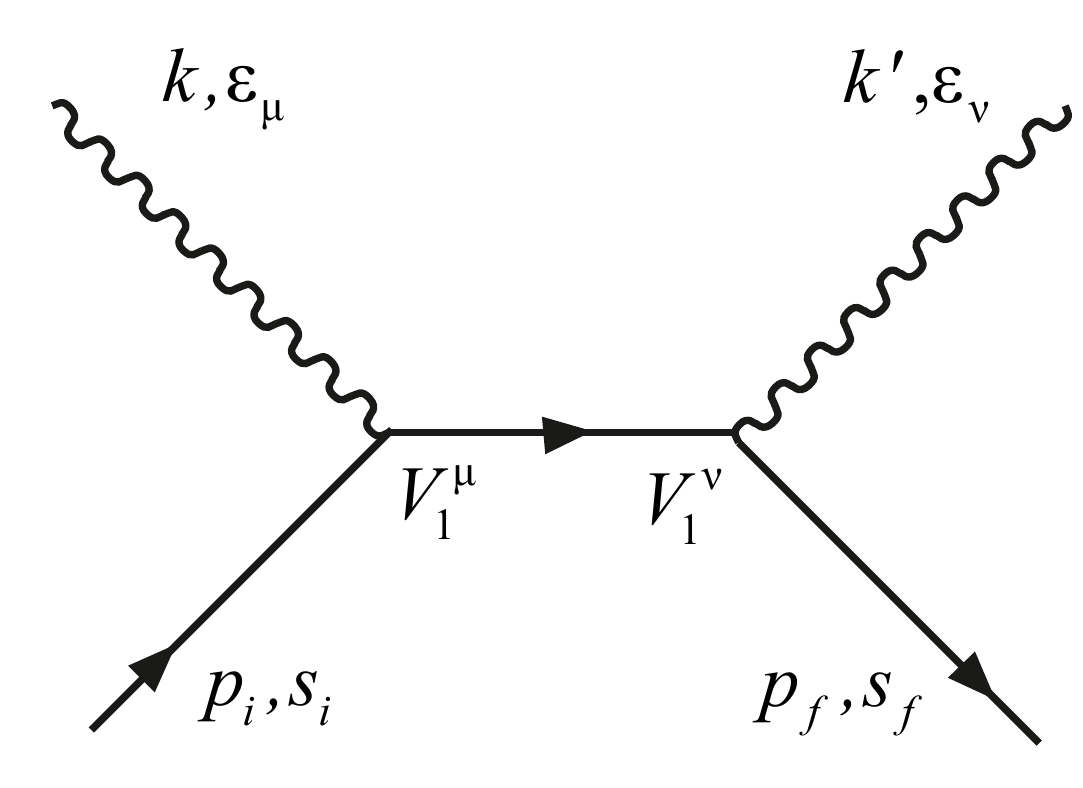} \\ (a)}
	\end{minipage}
	\hfill
	\begin{minipage}[h]{0.45\linewidth}
		\center{\includegraphics[width=0.85\linewidth]{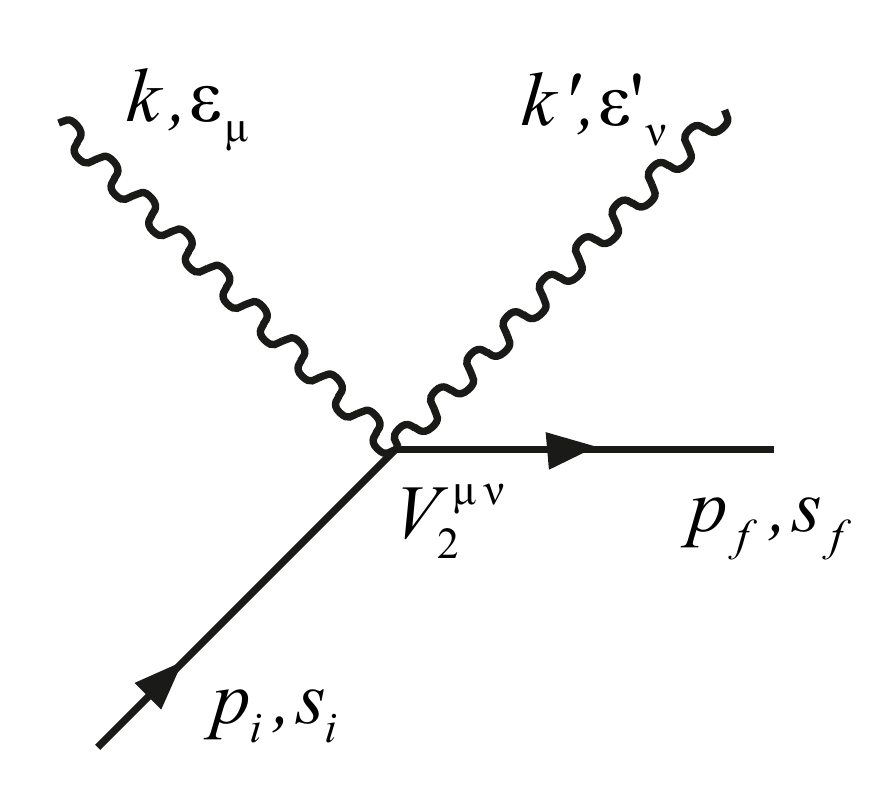} \\ (b)}
	\end{minipage}
	\vfill
	\begin{minipage}[h]{0.45\linewidth}
		\center{\includegraphics[width=1\linewidth]{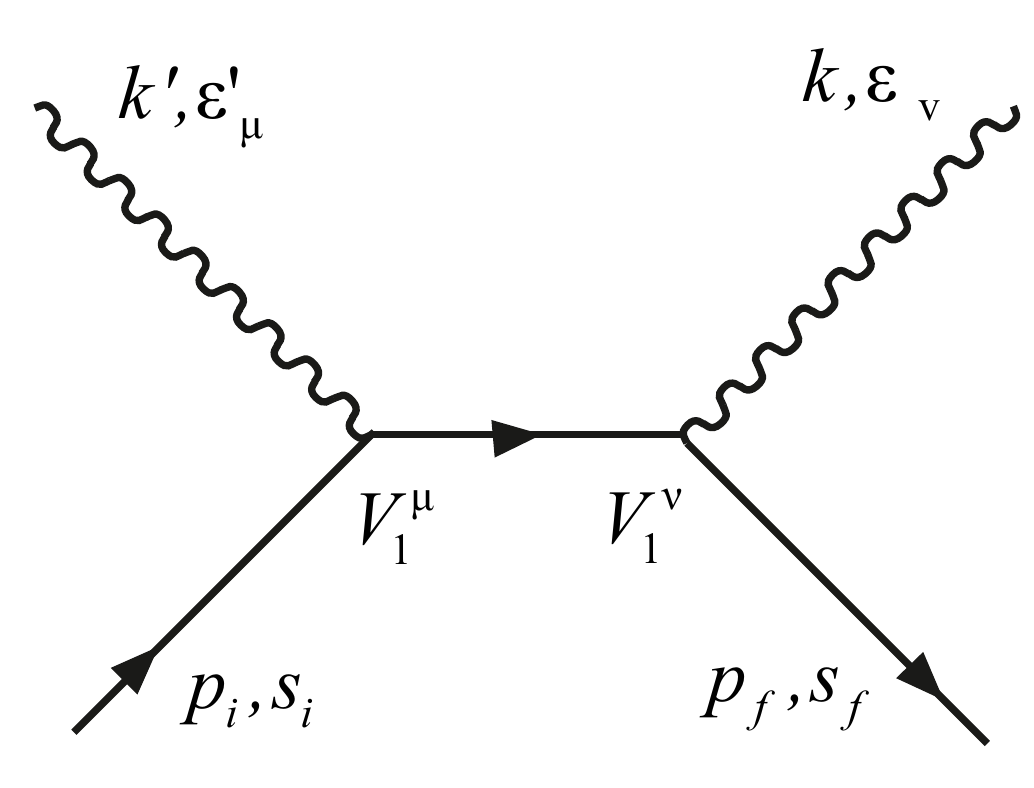} \\ (c)}
	\end{minipage}
	\hfill
	\begin{minipage}[h]{0.45\linewidth}
		\center{\includegraphics[width=0.85\linewidth]{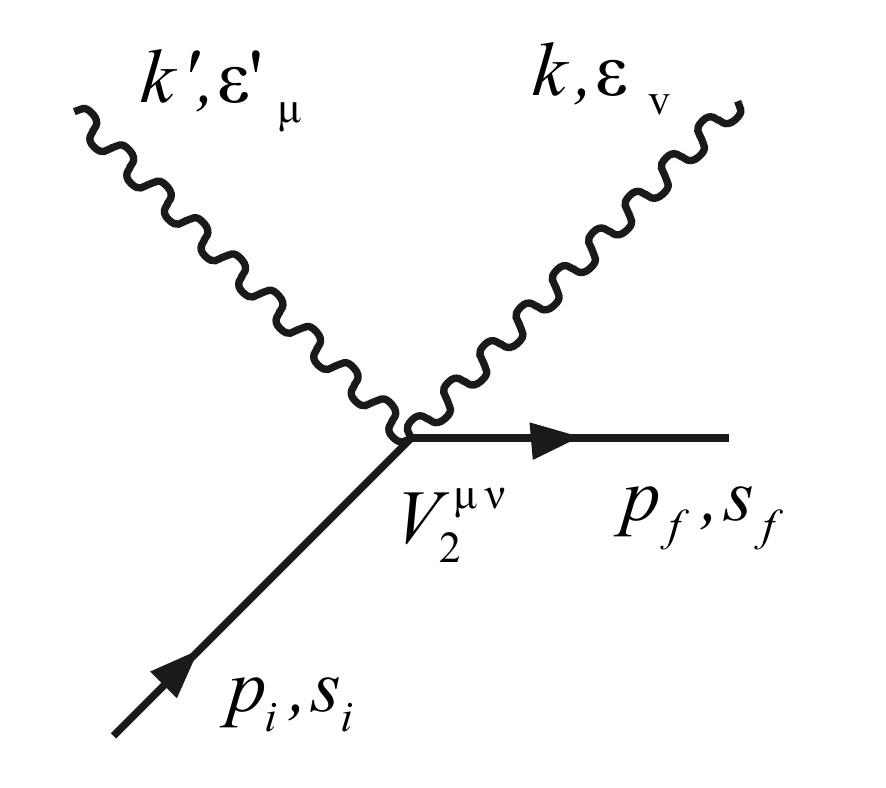} \\ (d)}
	\end{minipage}
	\caption{Compton scattering of electrons}
	\label{Fig.3}
\end{figure}


\begin{figure}[h!]
	\begin{minipage}[h]{0.5\linewidth}
		\center{\includegraphics[width=0.85\linewidth]{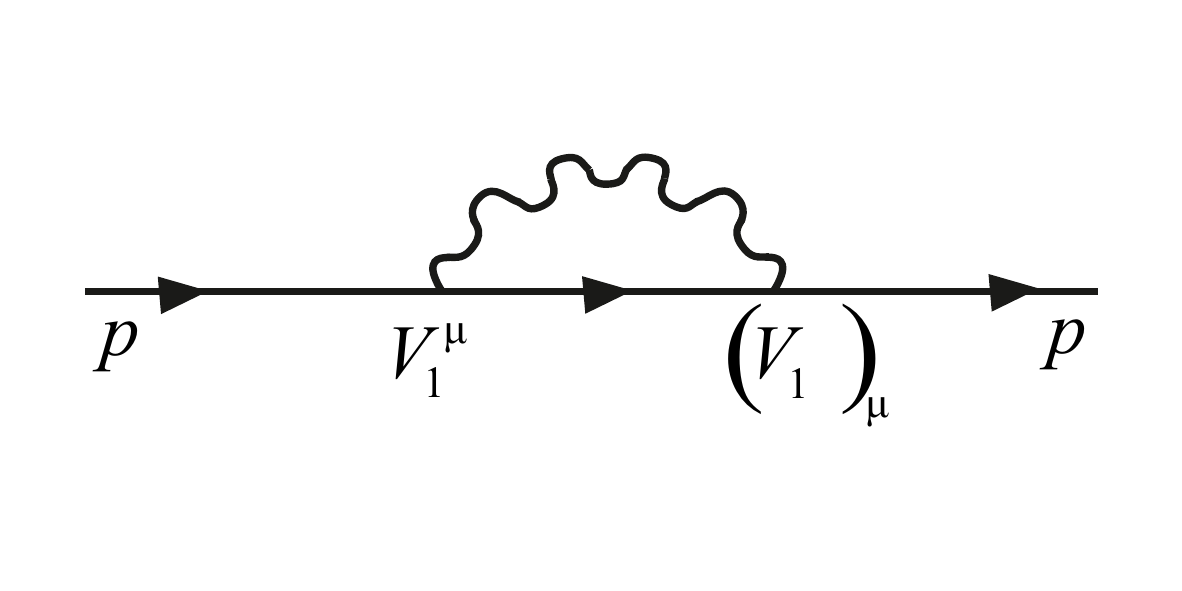} \\ (a)}
	\end{minipage}
	\hfill
	\begin{minipage}[h]{0.5\linewidth}
		\center{\includegraphics[width=0.90\linewidth]{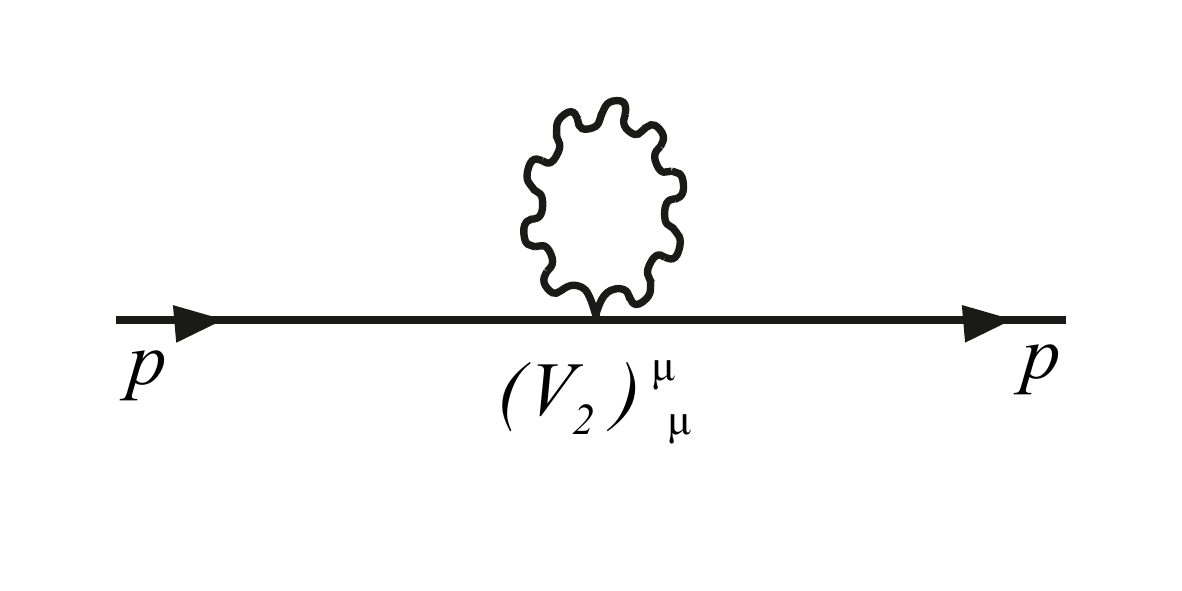} \\ (b)}
	\end{minipage}
	\caption{Self-energy of electron}
	\label{Fig.4}
\end{figure}

The calculations of self-energy with ultraviolet logarithmic divergence (see 
diagrams in Fig.4) also coincide with the results calculated in the Dirac 
representation. Some of the new nuances are discussed in the following 
section.

Fig. 5 presents diagrams for radiation corrections to electron scattering in 
the Coulomb field. Figures 5(a)--5(l) have to be taken into account when 
calculating the anomalous magnetic moment of an electron (see Sec. 5) and 
Lamb shift of atomic energy levels (see Sec. 6). In addition, at 
calculation of the Lamb shift, it is necessary to take into account Figs. 5(m)--5(o) related to the self-energy function of photon\footnote{ 
	The authors consciously avoid the phrase ''vacuum polarization'' by reasons 
	stated in Sec. 7.}
\footnote{In accordance with structure of interaction operator $V_3$ in Fig. 5(l), at first, a virtual photon is emitted from vertex, then the interaction with external electromagnetic fields is occurred and further a virtual photon is absorbed in vertex. All photon lines are crossed only in vertex of $(V_3)^{\mu \nu}_{\,\,\,\,\,\,\mu}$}. 


\begin{figure}[h!]
	\begin{minipage}[h]{0.28\linewidth}
		\center{\includegraphics[width=1\linewidth]{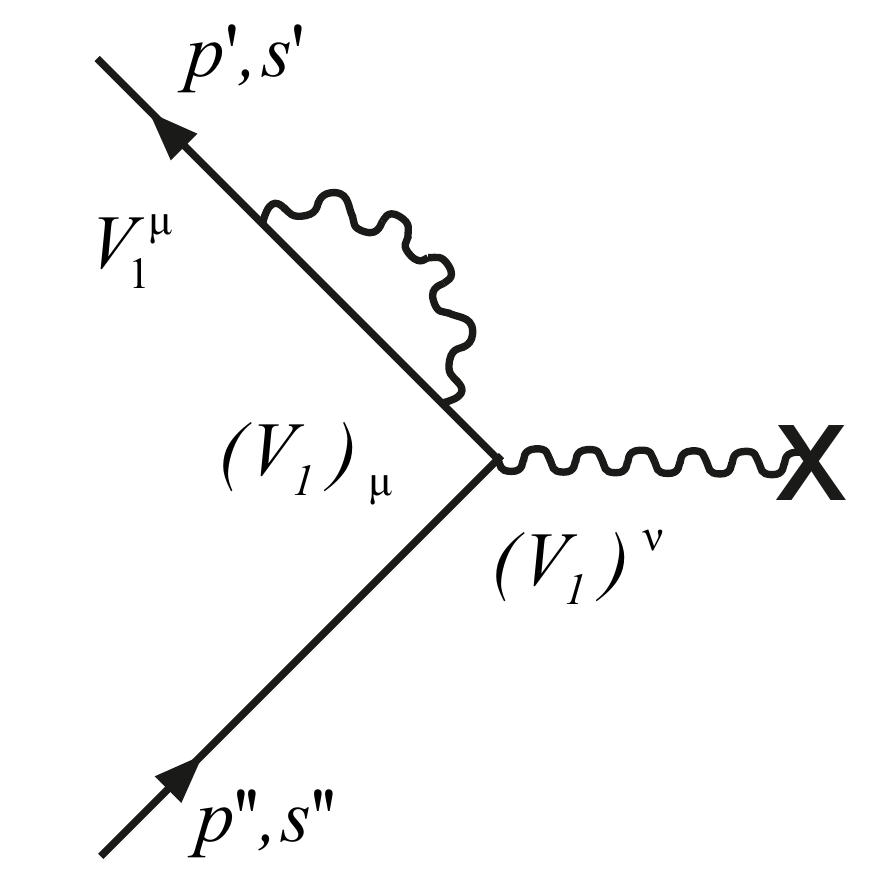} \\ (a)}
	\end{minipage}
	\hfill
	\begin{minipage}[h]{0.28\linewidth}
		\center{\includegraphics[width=1\linewidth]{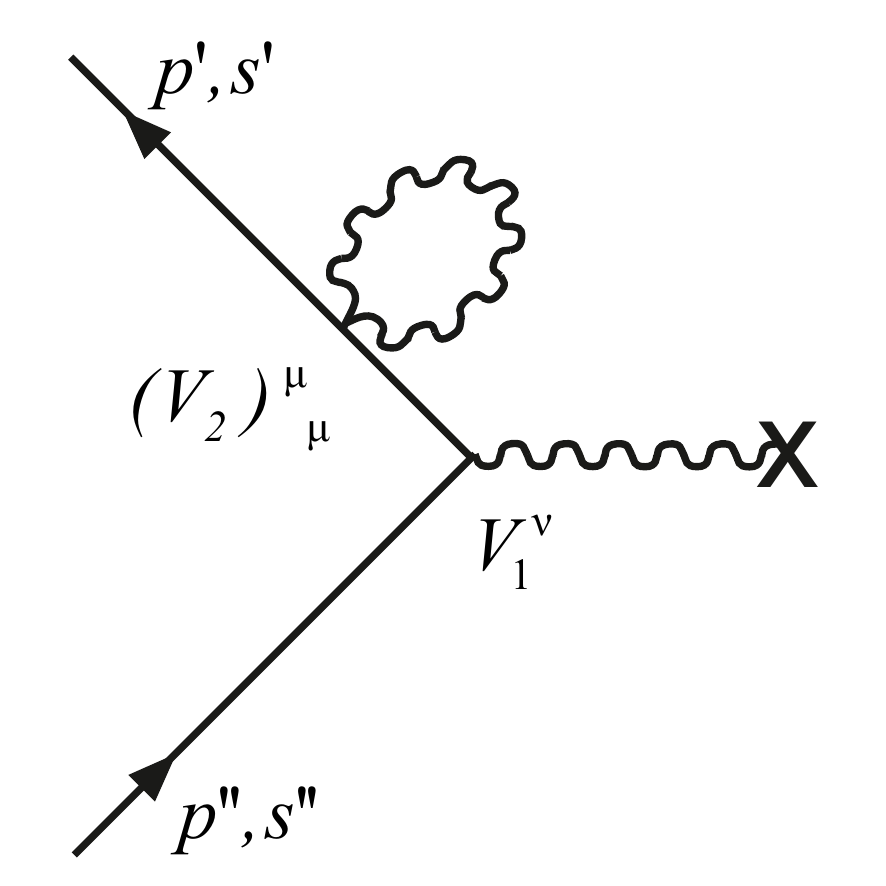} \\ (b)}
	\end{minipage}
	\hfill
	\begin{minipage}[h]{0.28\linewidth}
		\center{\includegraphics[width=1\linewidth]{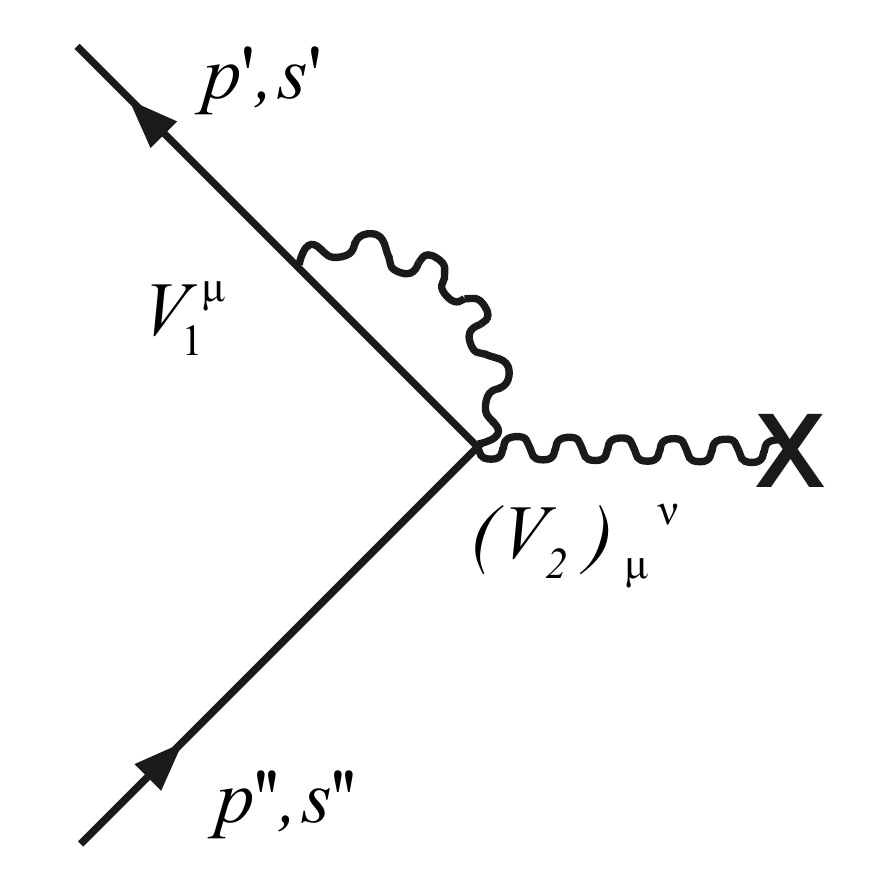} \\ (c)}
	\end{minipage}
	\vfill
	\begin{minipage}[h]{0.28\linewidth}
		\center{\includegraphics[width=1\linewidth]{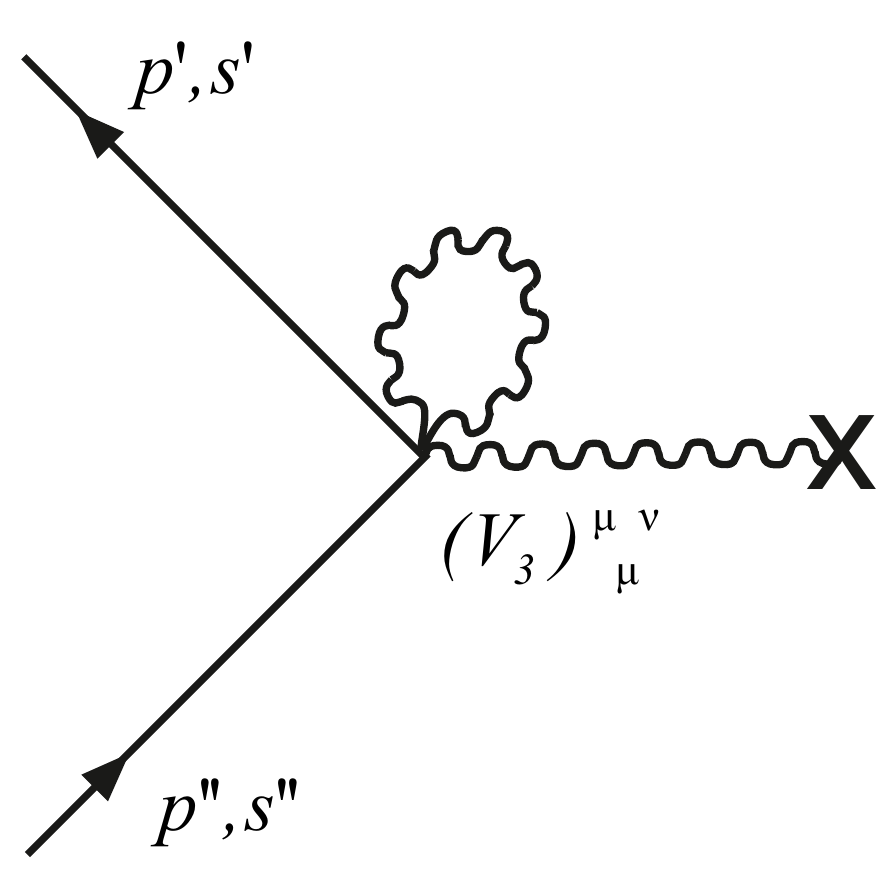} \\ (d)}
	\end{minipage}
	\hfill
	\begin{minipage}[h]{0.28\linewidth}
		\center{\includegraphics[width=1\linewidth]{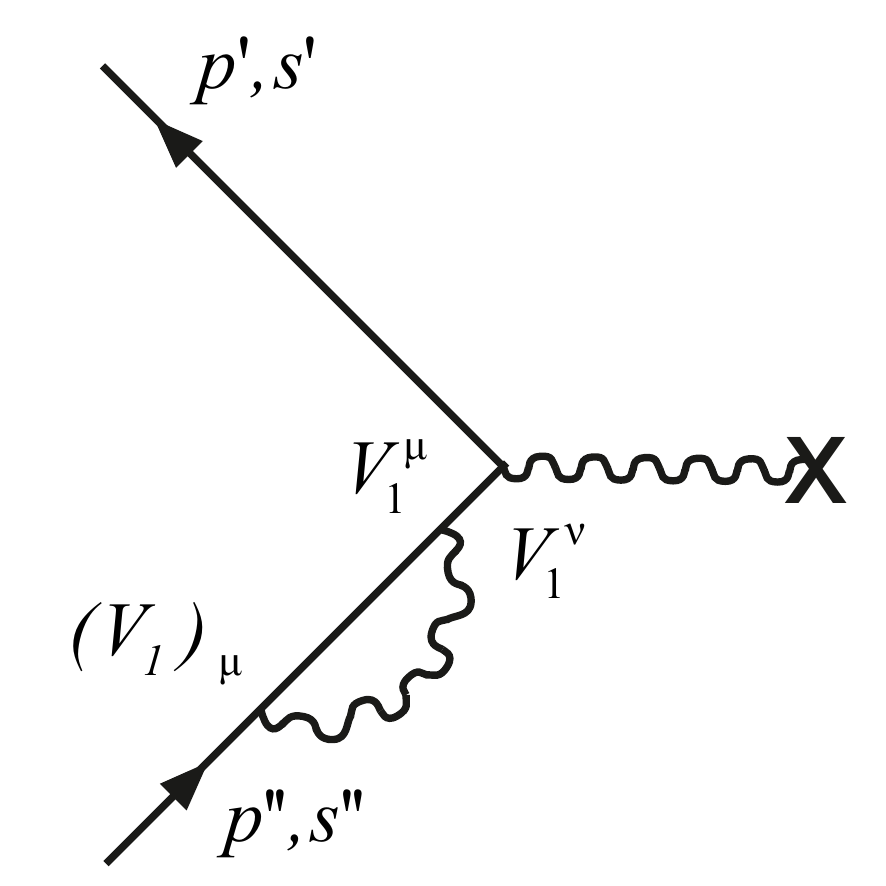} \\ (e)}
	\end{minipage}
	\hfill
	\begin{minipage}[h]{0.28\linewidth}
		\center{\includegraphics[width=1\linewidth]{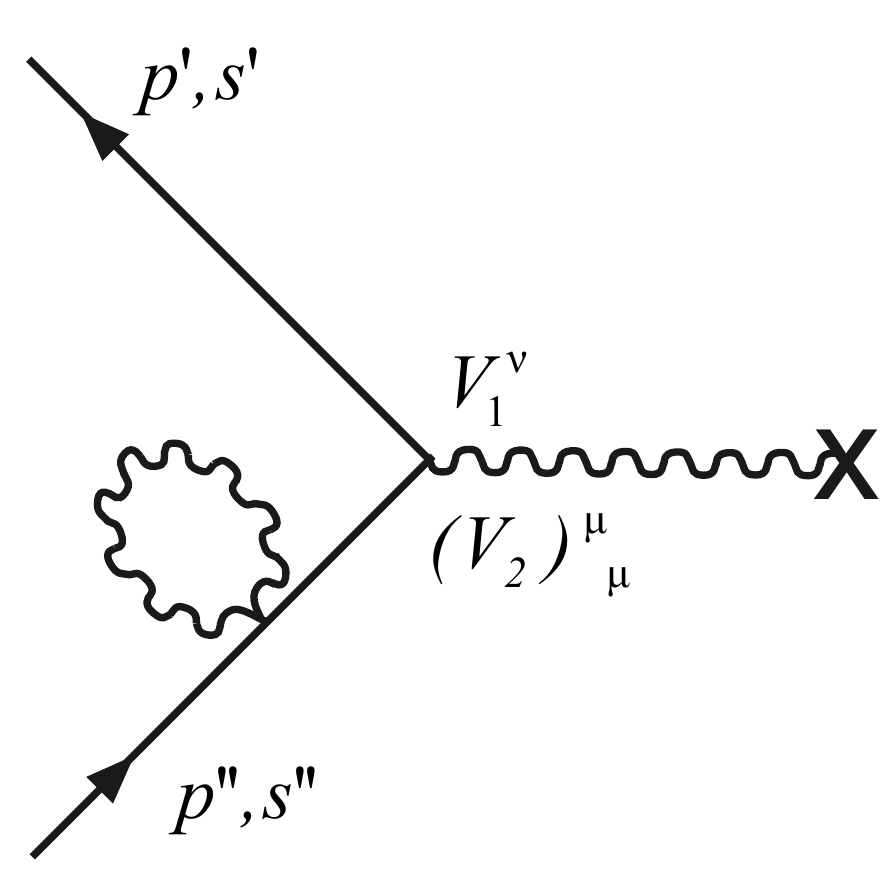} \\ (f)}
	\end{minipage}
	\vfill
	\begin{minipage}[h]{0.28\linewidth}
		\center{\includegraphics[width=1\linewidth]{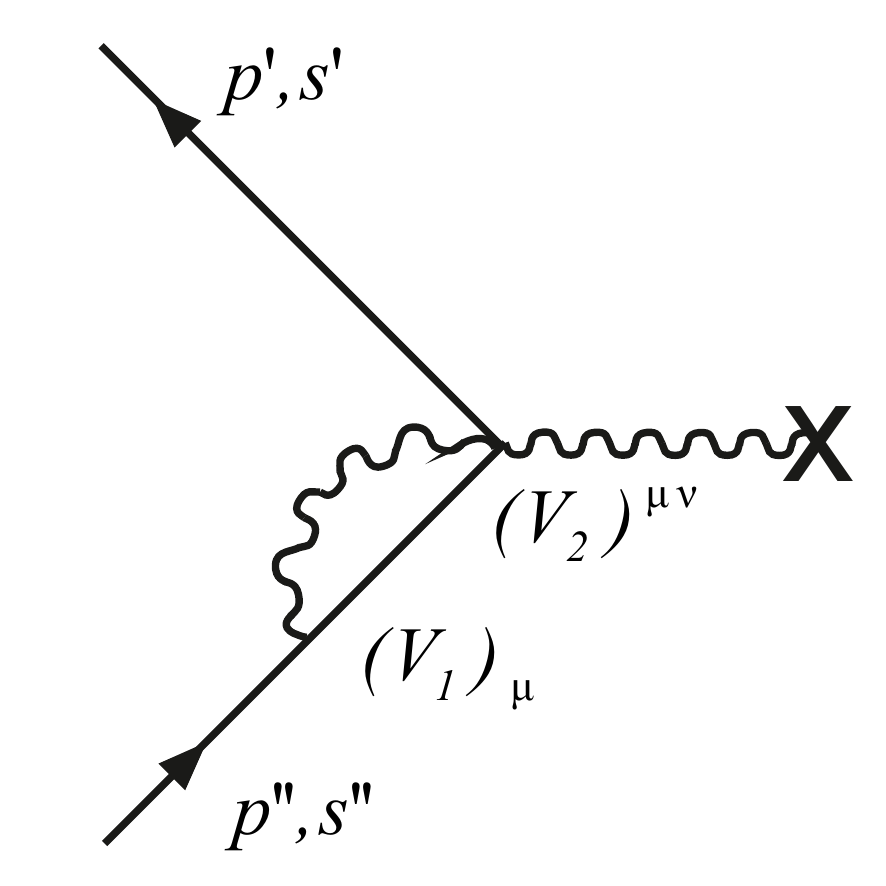} \\ (g)}
	\end{minipage}
	\hfill
	\begin{minipage}[h]{0.28\linewidth}
		\center{\includegraphics[width=1\linewidth]{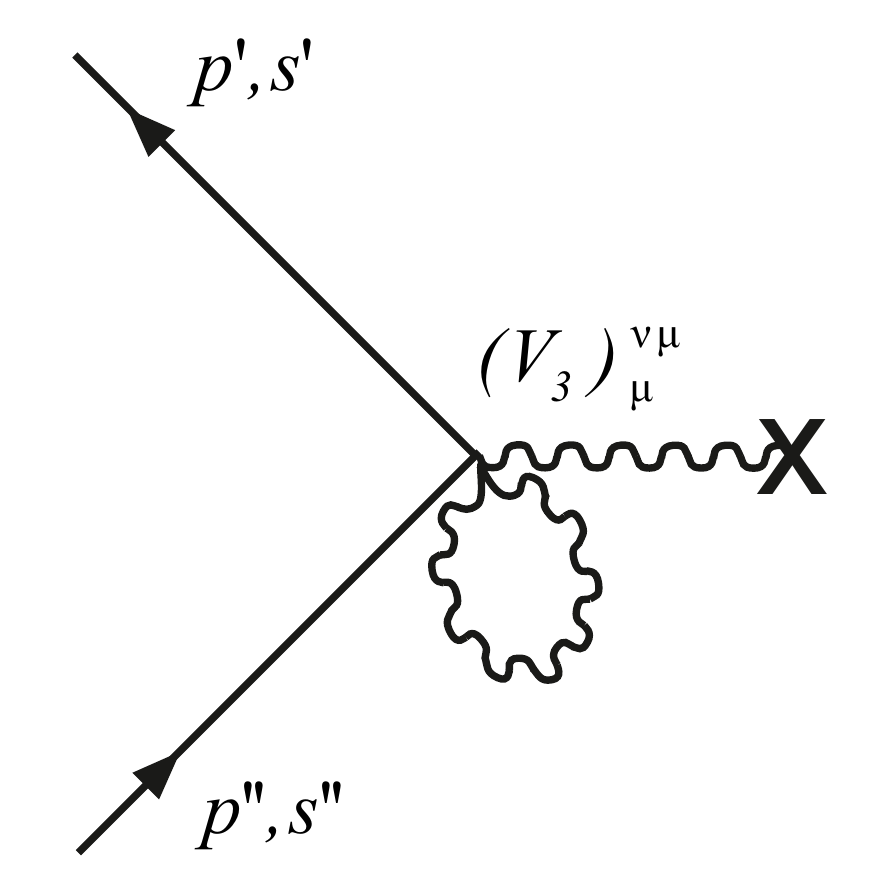} \\ (h)}
	\end{minipage}
	\hfill
	\begin{minipage}[h]{0.28\linewidth}
		\center{\includegraphics[width=1\linewidth]{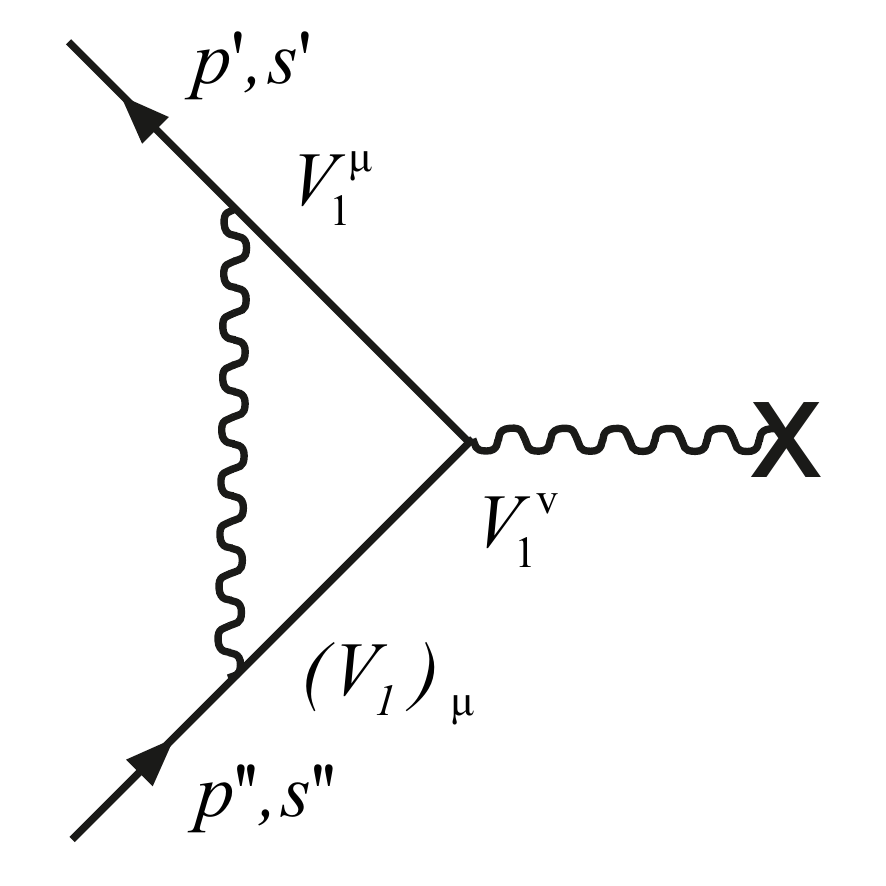} \\ (i)}
	\end{minipage}
	\vfill
	\begin{minipage}[h]{0.28\linewidth}
		\center{\includegraphics[width=1\linewidth]{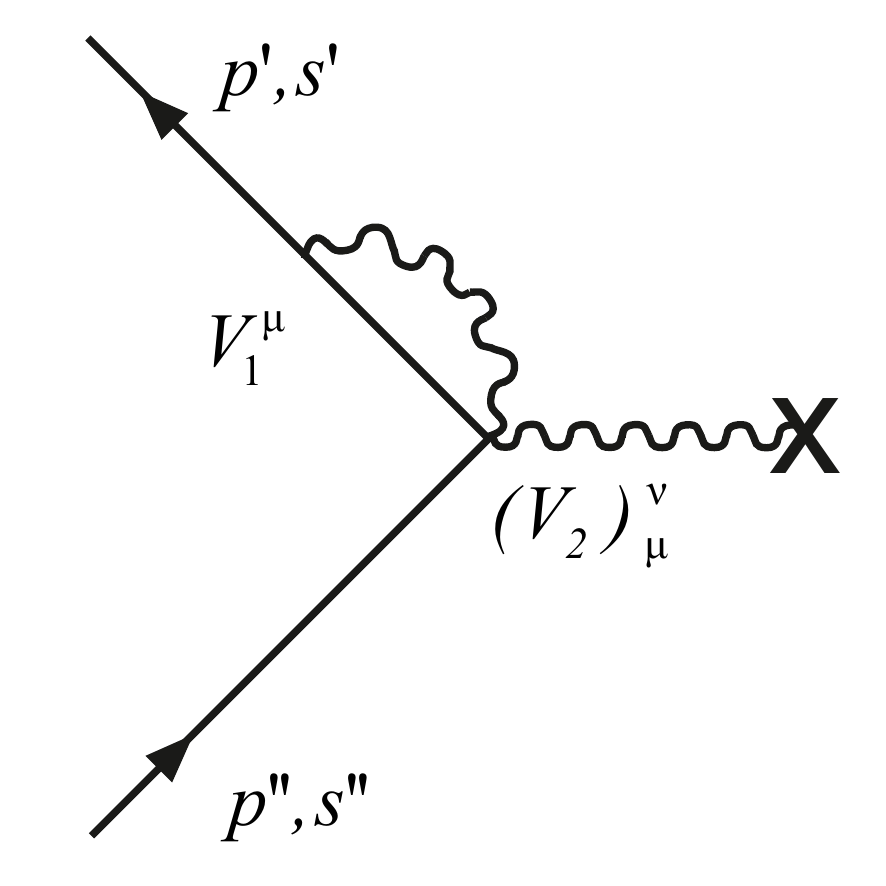} \\ (j)}
	\end{minipage}
	\hfill
	\begin{minipage}[h]{0.28\linewidth}
		\center{\includegraphics[width=1\linewidth]{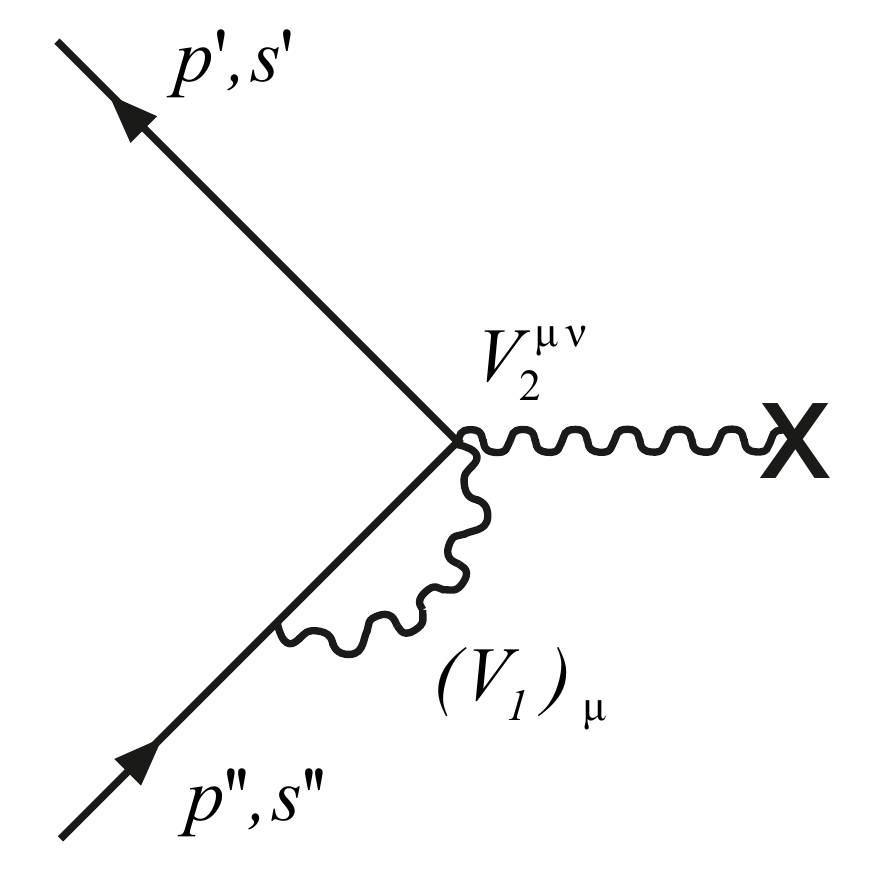} \\ (k)}
	\end{minipage}
	\hfill
	\begin{minipage}[h]{0.28\linewidth}
		\center{\includegraphics[width=1\linewidth]{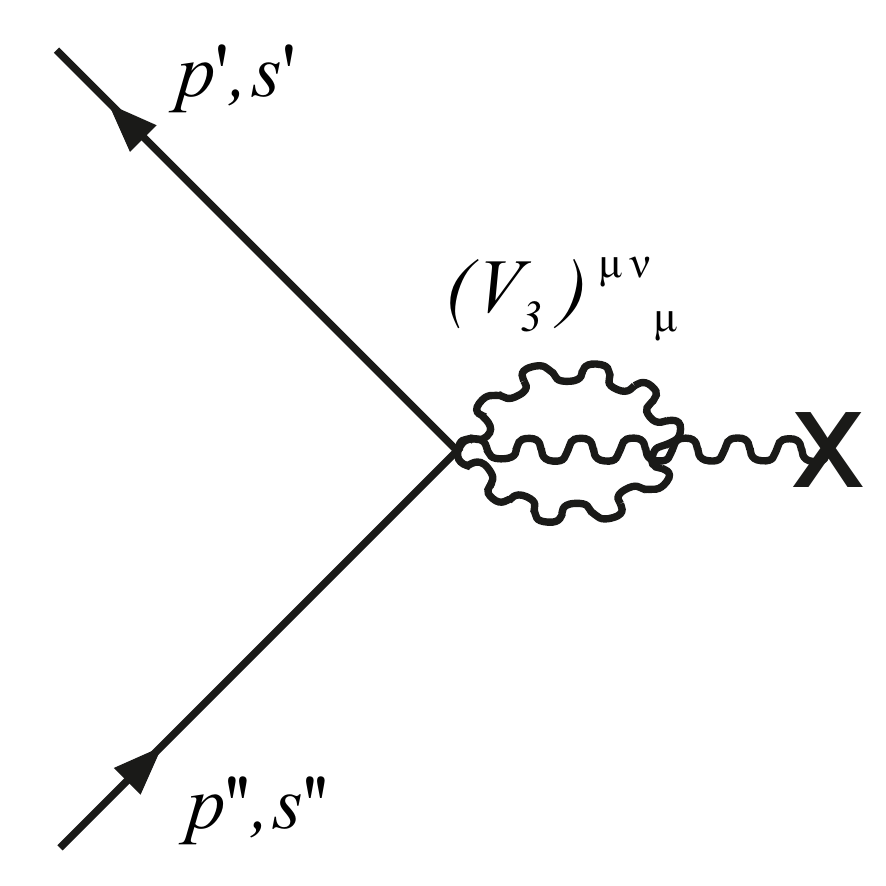} \\ (l)} 
	\end{minipage}
	\vfill
	\begin{minipage}[h]{0.28\linewidth}
		\center{\includegraphics[width=1\linewidth]{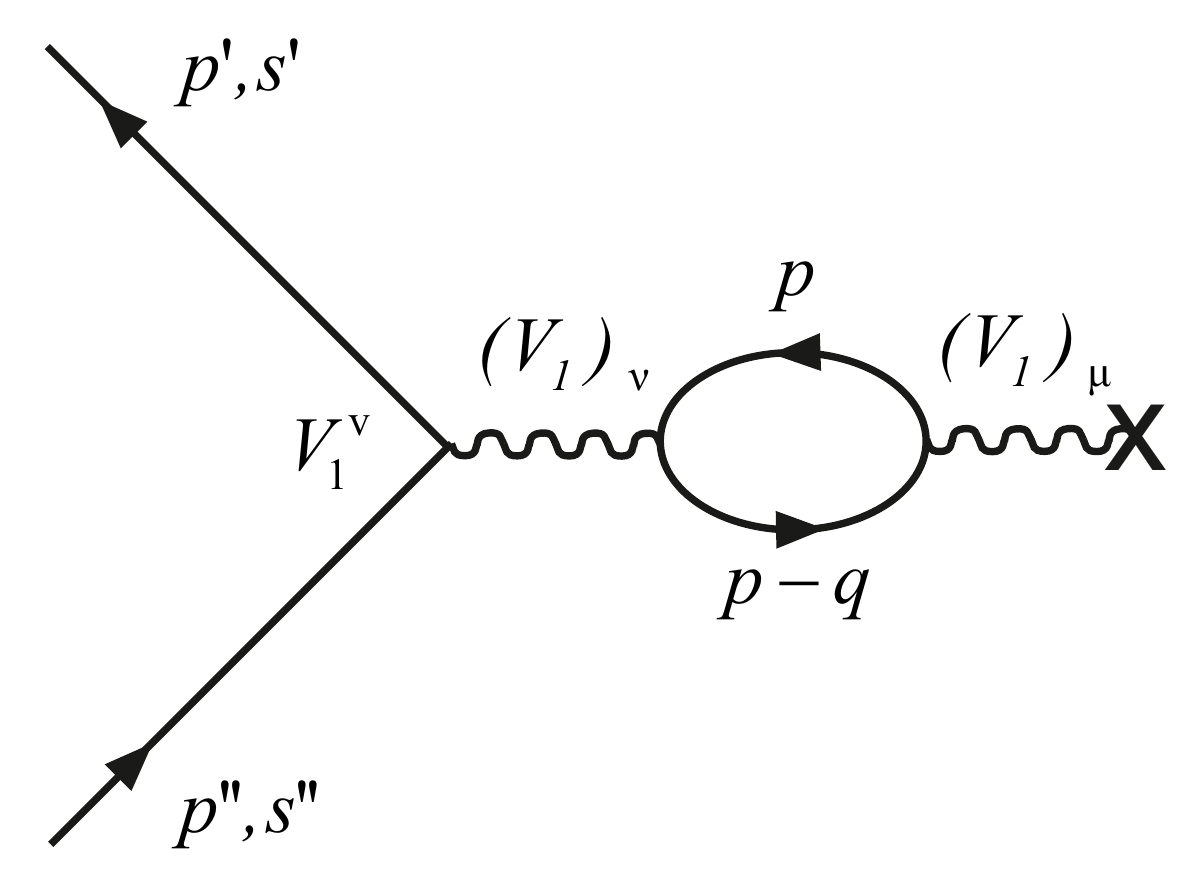} \\ (m)}
	\end{minipage}
	\hfill
	\begin{minipage}[h]{0.28\linewidth}
		\center{\includegraphics[width=1\linewidth]{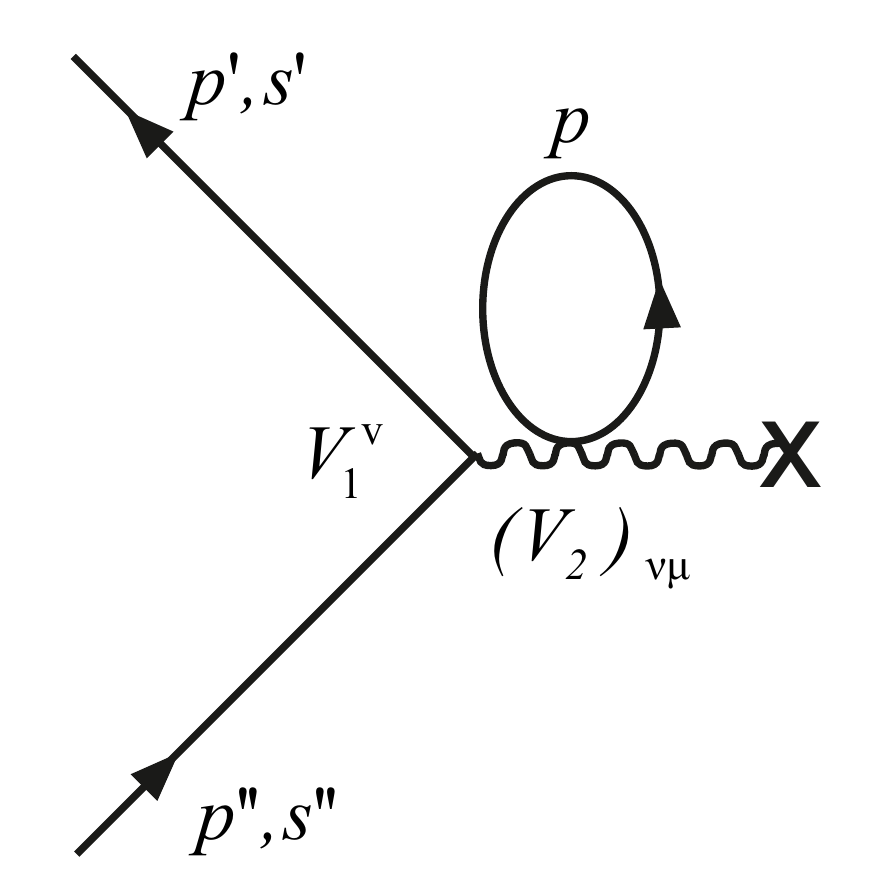} \\ (n)}
	\end{minipage}
	\hfill
	\begin{minipage}[h]{0.28\linewidth}
		\center{\includegraphics[width=1\linewidth]{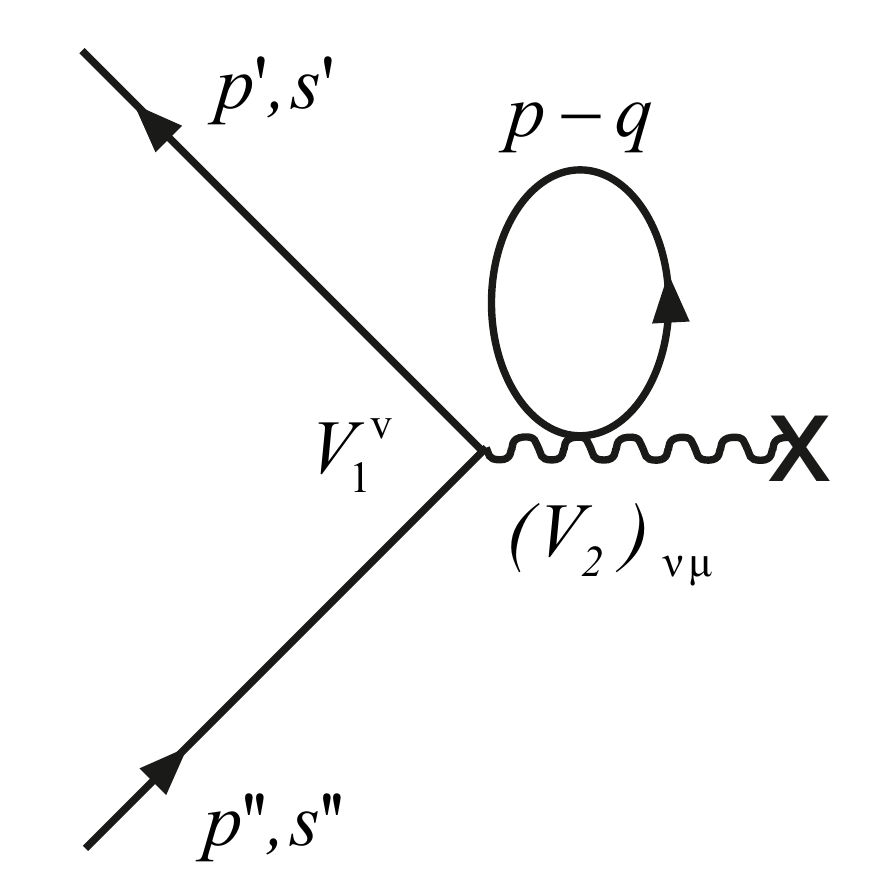} \\ (o)}
	\end{minipage}
	\caption{Radiation corrections to electron scattering in the external field.}
	\label{Fig.5}
\end{figure}

At implementation of renormalization of the electron mass, counterterms 
corresponding to Figs. 5(a)--5(d) with ${p}'=p^{IV}$ and Figs. 5(e)--5(h) with 
${{p}''}={{{p}'''}}_{\, }$ have to be subtracted. As a result, the 
contribution of Figs. 5(c), 5(d), and 5(g), 5(h) is zeroed. The contribution of 
renormalized Figs. 5(a), 5(b), and 5(e), 5(f) is finite and calculated in Secs. 5, 
6. The renormalization of the electron charge is the same as in the standard 
QED.

\section{Self-energy of an electron}

The Feynman diagrams to determine self-energy of an electron in the second 
order of the perturbation theory are presented in Fig.4.

Mass operator $-i\Sigma^{\left( 2 \right)} \left( p \right)$ is written 
as
\begin{equation}
	\label{eq36}
	-i\Sigma^{\left( 2 \right)} \left( p \right)=-\int {\frac{d^{4}k}{\left( 
			{2\pi } \right)^{4}k^{2}}\left[ {V_{1}^{\mu } \left( {{p}',{{{p}'''}}} 
			\right)\frac{1}{\left( {{{{p}'''}}} \right)^{\,2}-m^{2}}V_{1}^{\mu } \left( 
			{{{{p}'''}},{p}'} \right)+V_{2}^{\mu \mu } \left( {{p}',{{p}''}} \right)} 
		\right]} .
\end{equation}
Here, ${{{p}'''}}={p}'-k$; for the electron $I=\left( {\left( {{p}'} 
	\right)^{0\,\,}+m} \right)^{1/ 2}, \\ III=\left( {\left( {{p}'} 
	\right)^{0\,\,}-k^{0}+m} \right)^{1/2}$.
In compliance with (\ref{eq35})
\begin{equation}
	\label{eq37}
	\begin{array}{l}
		V_{1}^{0} \left( {{p}',{{{p}'''}}} \right)=e \left\lbrace \left( {{p}'} 
		\right)^{0\,\,}+\left( {{{{p}'''}}} \right)^{0\,\,}-\dfrac{\left( {{\rm {\bf 
						{p}'}}} \right)^{2}}{I\left( {I+III} \right)}+\dfrac{{\rm {\bm \sigma 
					\bf {p}' \bm\sigma {{{p}'''}}}}}{I\cdot III}-\dfrac{\left( {{\rm {\bf {{{p}'''}}}}} 
			\right)^{2}}{III\left( {I+III} \right)} \right\rbrace  \\ [12pt]
		=e \left\lbrace 2\left( {{p}'} \right)^{0\,\,}-k^{0}+\dfrac{2{\rm {\bf {p}'k}}}{III\left( 
			{I+III} \right)}-\dfrac{k^{2}}{III\left( {I+III} \right)}-\dfrac{{\rm {\bm 
					\sigma \bf {p}'\bm\sigma k}}}{I\cdot III} \right\rbrace, \\ 
	\end{array}
\end{equation}
\begin{equation}
	\label{eq38}
	V_{1}^{i} \left( {{p}',{{{p}'''}}} \right)=e \left\lbrace -\frac{III}{I}{\rm {\bm \sigma 
			\bf {p}'}}\sigma^{i}-\frac{I}{III}\sigma^{i}\left( {{\rm {\bm \sigma 
				\bf {p}'}}-{\rm {\bm \sigma \bf k}}} \right) \right\rbrace ,
\end{equation}
\begin{equation}
	\label{eq39}
	V_{1}^{0} \left( {{{{p}'''}},{p}'} \right)=e \left\lbrace 2\left( {{p}'} 
	\right)^{0\,\,}-k^{0}+\frac{2{\rm {\bf {p}'k}}}{III\left( {I+III} 
		\right)}-\frac{k^{2}}{III\left( {I+III} \right)}-\frac{{\rm {\bm \sigma 
				\bf k \bm\sigma {p}'}}}{I\cdot III} \right\rbrace ,
\end{equation}
\begin{equation}
	\label{eq40}
	V_{1}^{i} \left( {{{{p}'''}},{p}'} \right)=e \left\lbrace -\frac{I}{III}\left( {{\rm {\bm 
				\sigma \bf {p}'}}-{\rm {\bm \sigma \bf k}}} \right)\sigma^{i}-\frac{III}{I}\sigma 
	^{i}{\rm {\bm \sigma \bf {p}'}} \right\rbrace .
\end{equation}
In compliance with (A.1)
\begin{equation}
	\label{eq41}
	V_{2}^{00} =e^2 \left\lbrace -1+\frac{2{\rm {\bf {p}'k}}}{I\cdot III\left( {I+III} 
		\right)^{2}}+\frac{k^{2}}{III^{2}\left( {I+III} \right)^{2}} \right\rbrace ,
\end{equation}
\begin{equation}
	\label{eq42}
	V_{2}^{ik} =\frac{I^{2}}{III^{2}}\sigma^{i}\sigma^{k}.
\end{equation}
Upon substitution in (\ref{eq36}) of expressions (\ref{eq37}) - (\ref{eq42}), we obtain
\begin{equation}
	\label{eq43}
	-i\Sigma^{\left( 2 \right)} \left( p \right)=-4e^{2}\int 
	{\frac{d^{4}k}{\left( {2\pi } \right)^{4}k^{2}}\frac{pk+m^{2}}{\left( {p-k} 
			\right)^{2}-m^{2}}} .
\end{equation}
The expression (\ref{eq43}) agrees with the mass operator in Dirac representation. It is remarkable that 
expression (\ref{eq43}) with ultraviolet logarithmic divergence is obtained from the 
expression (\ref{eq36}) by using only virtual states with positive energy of the 
electron. In other words, expression (\ref{eq43}) is obtained by using only 
self-conjugated equation (\ref{eq12}) with positive energies of an electron. In 
Dirac representation it would lead to linear divergence of self-energy. In 
Dirac representation, only accounting of intermediate states with positive and negative 
energies leads to the view for self-energy with ultraviolet logarithmic 
divergence \cite{bib8}.

\section{Anomalous magnetic moment of an electron}

The Feynman diagrams, necessary to calculate corrections to the magnetic 
moment of an electron in the second order of the perturbation theory, are 
given in Fig.5.

In the process under consideration, the static potentials are $A^{i}\left( 
{{\rm {\bf x}}} \right)\ne 0, \\ A^{0}\left( {{\rm {\bf x}}} \right)=0$. The 
amplitude in the first order of the perturbation theory is 
\begin{equation}
	\label{eq44}
	-e\frac{i\delta \left( {E_{f} -E_{i} } \right)}{\left( {2\pi } 
		\right)^{2}2E_{i} }\bar{{U}}_{s_{f} } \left\langle {{\rm {\bf p}}_{f} \left| 
		{-{\rm {\bm \sigma \bf p}}_{f} {\rm {\bm \sigma \bf A}}-{\rm {\bm \sigma \bf A \bm \sigma 
					p}}_{i} } \right|{\rm {\bf p}}_{i} } \right\rangle >U_{s_{i} } .
\end{equation}
Let us denote in (\ref{eq44})
\begin{equation}
	\label{eq45}
	L={\rm {\bm \sigma \bf p}}_{f} {\rm {\bm \sigma \bf A}}+{\rm {\bm \sigma \bf A\bm\sigma 
			p}}_{i} ={\rm {\bf p}}_{f} {\rm {\bf A}}+{\rm {\bf Ap}}_{i} +{\rm {\bm 
			\sigma  \bf H}}.
\end{equation}
It what fallows, we will consider electron motion in the weak magnetic field and take 
into account only addends proportional to $p^{i}$ and ${\rm {\bm \sigma 
		\bf H}}$, where ${\rm {\bf H}}$ is a magnetic field. In our denotations, 
\begin{equation*}
\begin{array}{l}
{\rm {\bf p}}_{f} ={\rm {\bf {p}'}},\,\,{\rm {\bf p}}_{i} ={\rm {\bf 
		{{p}''}}},\,\,{\rm {\bf {{{p}'''}}}}={\rm {\bf {p}'}}-{\rm {\bf 
		k}},\,\,\,{\rm {\bf p}}^{IV}={\rm {\bf {{p}''}}}-{\rm {\bf k}},
\quad \\ [10pt]
q={p}'-{{p}''}=\left\{ {\begin{array}{l}
		q^{0}=0,\,\,\left( {{p}'} \right)^{0}=\left( {{{p}''}} \right)^{0}, \\ 
		q^{i}=\left( {{p}'} \right)\,^{i}-\left( {{{p}''}} \right)^{\,i}. \\ 
\end{array}} \right.	
\end{array}
\end{equation*}

Let us consider the contribution of Figs. 5(i)--5(l) to the anomalous 
magnetic moment. For diagrams of this type, the equation $III=IV$ holds true. 
Let us write down the amplitude as
\begin{equation}
	\label{eq46}
	-\frac{i\delta \left( {{E}'-{{E}''}} \right)}{\left( {2\pi } 
		\right)^{2}2{{E}''}}\bar{{U}}_{{s}'} \left\langle {{\rm {\bf {p}'}}\left| 
		{\frac{e^{2}}{\left( {2\pi } \right)^{4}}\int {\frac{d^{4}k}{k^{2}}\Lambda 
				^{i}\left( {{\rm {\bf {p}'}},{\rm {\bf {{p}''}}}} \right)A_{{\rm {\bf 
							{p}'}}-{\rm {\bf {{p}''}}}}^{i} } } \right|{\rm {\bf {{p}''}}}} 
	\right\rangle >U_{{{s}'}'} .
\end{equation}

\subsection{Contribution of Fig. 5(i)}

\begin{equation}
	\label{eq47}
	\begin{array}{l}
	\left( {\Lambda^{i}A^{i}} \right)_{5i} =\left[ {V_{1}^{0} \left( 
			{{p}',{{{p}'''}}} \right)} \right.V_{1}^{i} \left( {{{{p}'''}},p^{IV}} 
		\right)A^{i}V_{1}^{0} \left( {p^{IV},{{p}''}} \right) \\ [12pt]
		\left. {-V_{1}^{l} \left( {{p}',{{{p}'''}}} \right)V_{1}^{i} \left( 
			{{{{p}'''}},p^{IV}} \right)A^{i}V_{1}^{l} \left( {p^{IV},{{p}''}} \right)} 
		\right]\dfrac{1}{\left( {\left( {{{{p}'''}}} \right)^{2}-m^{2}} \right)\left( 
			{\left( {p^{IV}} \right)^{2}-m^{2}} \right)} \\ [12pt]
		= 	e^3 \left\{ {-\left( {2m-k^{0}-\dfrac{{\rm {\bf k}}^{2}}{III\left( {I+III} 
					\right)}} \right)^{2}L }\right. \\ [12pt]
		\left. { -\left[ {\left( {2m-k^{0}-\dfrac{{\rm {\bf 
								k}}^{2}}{III\left( {I+III} \right)}} \right)\dfrac{1}{I\cdot III}+1} \right]
2{\rm {\bf kA}}\left( {{\rm {\bm \sigma \bf {p}'\bm\sigma k}}+{\rm 
					{\bm \sigma \bf k \bm \sigma {{p}''}}}} \right) } \right. \\ [12pt]
	\left. { -\dfrac{I^{2}}{III^{2}}{\rm {\bf 
					k}}^{2}L+\dfrac{I^{2}}{III^{2}}\left[ {\left( {2{\rm {\bf {p}'k}}} 
				\right){\rm {\bm \sigma \bf k\bm\sigma \bf A}}+\left( {2{\rm {\bf {{p}''}k}}} 
				\right){\rm {\bm \sigma \bf A\bm\sigma \bf k}}} \right]} \right\} \\ [12pt]
		\times \dfrac{1}{\left[ {\left( {2m-k^{0}} \right)k^{0}+{\rm {\bf k}}^{2}} 
			\right]^{2}}. \\ 
	\end{array}
\end{equation}

\subsection{Contribution of Figs. 5(j) and 5(k)}

\begin{equation}
	\label{eq48}
	\begin{array}{l}
		\left( {\Lambda^{i}A^{i}} \right)_{5j} +\left( {\Lambda^{i}A^{i}} 
		\right)_{5k} =\left[ {V_{1}^{0} \left( {{p}',{{{p}'''}}} \right)} 
		\right.V_{2}^{i0} \left( {{{{p}'''}},{{p}'}'} \right)A^{i} \\ [12pt]
		\left. {-V_{1}^{l} \left( {{p}',{{{p}'''}}} \right)V_{2}^{il} \left( 
			{{{{p}'''}},{{p}''}} \right)A^{i}} \right]\dfrac{1}{\left( {{{{p}'''}}} 
			\right)^{2}-m^{2}}+\left[ {V_{2}^{0i} \left( {{p}',p^{IV}} \right)} 
		\right.A^{i}V_{1}^{\,0} \left( {p^{IV},{{p}''}} \right) \\ [12pt]
		-\left. {V_{2}^{li} \left( {{p}',p^{IV}} \right)A^{i}V_{1}^{l} \left( 
			{p^{IV},{{p}''}} \right)} \right]\dfrac{1}{\left( {p^{IV}} 
			\right)^{2}-m^{2}} \\ [12pt]
		= 	e^3 \left\{ {\left( {2m-k^{0}-\dfrac{{\rm {\bf k}}^{2}}{III\left( {I+III} 
					\right)}} \right)\left( {\dfrac{2}{III\left( {I+III} \right)}-\dfrac{1}{I\cdot 
					III}} \right)L-L-\dfrac{I^{2}}{III^{2}}} \right. \\ [12pt]
		\left. {+\dfrac{1}{I\cdot III^{2}\left( {I+III} \right)}2{\rm {\bf 
					kA}}\left( {{\rm {\bm \sigma \bf {p}'\bm\sigma \bf k}}+{\rm {\bm \sigma \bf k\bm\sigma 
						\bf {{p}''}}}} \right)} \right\} \dfrac{1}{\left[ {-\left( {2m-k^{0}} 
				\right)k^{0}-{\rm {\bf k}}^{2}} \right]} \\ [12pt]
		-	e^3  \dfrac{I^{2}}{III^{2}}\left[ {2{\rm {\bf {p}'k\bm\sigma k\bm\sigma A}}+2{\rm {\bf 
					{{p}''}k\bm\sigma A\bm\sigma k}}} \right]\dfrac{1}{\left[ {\left( {2m-k^{0}} 
				\right)k^{0}+{\rm {\bf k}}^{2}} \right]^{2}}. \\ 
	\end{array}
\end{equation}

\subsection{Contribution of Fig. 5(l)}

\begin{equation}
	\label{eq49}
	\left( {\Lambda^{i}A^{i}} \right)_{5l} =V_{3}^{0i0} A^{i}=\frac{e^3 L}{I\cdot 
		III\left( {I+III} \right)^{2}}.
\end{equation}
The sum of expressions (\ref{eq47}) - (\ref{eq49}) is
\begin{equation}
	\label{eq50}
	\begin{array}{l}
	\left( {\Lambda^{i}A^{i}} \right)_{5i+5j+5k+5l}  \\ [10pt]
	= -e^3\,\,\dfrac{\left( {2\left( 
			{k^{0}} \right)^{2}-2{\rm {\bf k}}^{2}-8mk^{0}+4m^{2}} \right)L+4{\rm {\bf 
				kA}}\left( {{\rm {\bm \sigma \bf {p}'\bm\sigma \bf k}}+{\rm {\bm \sigma \bf k\bm\sigma 
					\bf {{p}''}}}} \right)}{\left[ {\left( {2m-k^{0}} \right)k^{0}+{\rm {\bf 
					k}}^{2}} \right]^{2}}.
\end{array}
\end{equation}
The expression (\ref{eq50}) coincides with the contribution of single diagram in Fig. 5(i) with vertices $-ie\gamma^{\mu }$ in Dirac representation (see, for example, Ref. \cite{bib8}).

Let us turn to self-energy diagrams 5(a)--5(h). For each diagrams, we will 
subtract the appropriate mass counterterm connected with renormalization of 
electron mass (see Secs. 29, 30 in W.Heitler monograph \cite{bib8}).

For Figs. 5(a)--5(d), equalities $I=II=IV$ hold true. For Figs. 5(e)--5(h) 
equalities $I=II=III$ hold true. If singular denominators do not appear in 
calculation of contribution from diagrams, then these contributions are 
completely compensated by appropriate mass counterterms. At appearance of 
singular denominators, we will use the Heitler limiting process \cite{bib8}. In these cases, the convergent expressions are 
remained at subtraction of mass counterterms. In these cases subtractions of mass counterterms lead to finite expressions.

\subsection{Contribution of Figs. 5(c) and 5(d)}

\begin{equation}
	\label{eq51}
	\begin{array}{l}
		\left( {\Lambda^{i}A^{i}} \right)_{5c+5d} =\left[ {V_{1}^{0} \left( 
			{{p}',{{{p}'''}}} \right)V_{2}^{0i} \left( {{{{p}'''}},{{p}''}} 
			\right)A^{i}-V_{1}^{l} \left( {{p}',{{{p}'''}}} \right)V_{2}^{li} \left( 
			{{{{p}'''}},{{p}''}} \right)A^{i}} \right]\dfrac{1}{\left( {{{{p}'''}}} 
			\right)^{\,2}-m^{2}} \\ [12pt]
		+V_{3}^{00i} A^{i}= e^3 \left[ {\left( {2m-k^{0}-\dfrac{{\rm {\bf 
							k}}^{2}}{III\left( {I+III} \right)}} \right)\left( {\dfrac{L}{I\left( {I+III} 
					\right)}-\dfrac{{\rm {\bm \sigma \bf {p}'\bm\sigma \bf A}}}{I\cdot III}} \right)-} 
		\right.\dfrac{{\rm {\bf k}}^{2}}{I^{2}III^{2}}{\rm {\bm \sigma \bf {p}'\bm\sigma 
				\bf A}} \\ [12pt]
		+\dfrac{3III^{2}}{I^{2}}{\rm {\bm \sigma \bf{p}'\bm\sigma \bf A}}-{\rm {\bm \sigma 
				\bf {p}'\bm\sigma \bf A}}+\left( {2m-k^{0}-\dfrac{{\rm {\bf k}}^{2}}{III\left( {I+III} 
				\right)}} \right)\dfrac{{\rm {\bm \sigma \bf k\bm\sigma \bf A}}}{I\cdot III}+{\rm {\bm 
				\sigma \bf k \bm\sigma \bf A}} \\ [12pt]
		\left. {+\dfrac{2{\rm {\bf {p}'k}}}{I\cdot III^{2}\left( {I+III} 
				\right)}{\rm {\bm \sigma \bf k\bm\sigma \bf A}}} \right]\dfrac{1}{\left( {{{{p}'''}}} 
			\right)^{\,2}-m^{2}}+\dfrac{e^3}{2I^{2}\left( {I+III} \right)^{2}}L \\ [12pt]
		-\dfrac{e^3}{I^{2}\cdot III\left( {I+III} \right)}{\rm {\bm \sigma \bf {p}'\bm\sigma 
				\bf A}}-\dfrac{e^3}{I\cdot III^{2}\left( {I+III} \right)}{\rm {\bm \sigma \bf k\bm\sigma 
				\bf A}}. \\ 
	\end{array}
\end{equation}

\subsection{Contribution of Figs. 5(g) and 5(h)}

\begin{equation}
	\label{eq52}
	\begin{array}{l}
		\left( {\Lambda^{i}A^{i}} \right)_{5g+5h} =\left[ {V_{2}^{i0} \left( 
			{{p}',p^{IV}} \right)A^{i}V_{1}^{0} \left( {p^{IV},{{p}''}} 
			\right)-V_{2}^{il} \left( {{p}',p^{IV}} \right)A^{i}V_{1}^{l} \left( 
			{p^{IV},{{p}''}} \right)} \right] \\ [12pt]
		\times \dfrac{1}{\left( {p^{IV}} \right)^{\,2}-m^{2}}+V_{3}^{i\,00} 
		A^{i}=e^3 \left[ {\left( {2m-k^{0}-\dfrac{{\rm {\bf k}}^{2}}{IV\left( {I+IV} 
					\right)}} \right)} \right.\left( {\dfrac{L}{I\left( {I+IV} 
				\right)}-\dfrac{{\rm {\bm \sigma \bf A\bm\sigma \bf {{p}''}}}}{I\cdot IV}} \right) \\ [12pt]
		-\dfrac{{\rm {\bf k}}^{2}}{IV\left( {I+IV} \right)}{\rm {\bm \sigma \bf A \bm\sigma 
				\bf {{p}''}}}+\left( {\dfrac{3IV^{2}}{I^{2}}-1} \right){\rm {\bm \sigma \bf A\bm\sigma 
				\bf {{p}''}}}+\left( {2m-k^{0}-\dfrac{{\rm {\bf k}}^{2}}{IV\left( {I+IV} 
				\right)}} \right)\dfrac{{\rm {\bm \sigma \bf A\bm\sigma \bf k}}}{I\cdot IV} \\ [12pt]
		\left. {+{\rm {\bm \sigma \bf A\bm\sigma \bf k}}+\dfrac{2{\rm {\bf {{p}''}k}}}{I\cdot 
				IV^{2}\left( {I+IV} \right)}{\rm {\bm \sigma \bf A \bm\sigma \bf k}}} 
		\right]\dfrac{1}{\left( {p^{IV}} \right)^{\,2}-m^{2}}+\dfrac{e^3}{2I^{2}\left( 
			{I+IV} \right)^{2}}L \\ [12pt]
		-\dfrac{e^3}{I^{2}\cdot IV\left( {I+IV} \right)}{\rm {\bm \sigma \bf A\bm\sigma 
				\bf {{p}''}}}-e^3 \dfrac{{\rm {\bm \sigma \bf A\bm\sigma \bf k}}}{I\cdot IV^{2}\left( {I+IV} 
			\right)}. \\ 
	\end{array}
\end{equation}
Here,  $IV=III$, since $p_{0}^{IV} ={p}_{0}'' -k_{0} 
={p}_{0}''' ={p}'_{0} -k_{0} $.
Taking this into account, the total contribution of (\ref{eq51}), (\ref{eq52}) is
\begin{equation}
	\label{eq53}
	\begin{array}{l}
		\left( {\Lambda^{i}A^{i}} \right)_{5c+5d+5g+5h} =e^3 \left[ {-\dfrac{{\rm {\bf 
						k}}^{2}IV\left( {2I+III} \right)}{I^{2}III\left( {I+III} 
				\right)^{2}}+\dfrac{3III^{2}}{I^{2}}-1} \right] \\ [12pt]
			\times \dfrac{L}{\left( {-\left( 
				{2m-k^{0}} \right)k^{0}-{\rm {\bf k}}^{2}} \right)}-e^3 \dfrac{4\left( {{\rm {\bf {p}'k\bm\sigma \bf k\bm\sigma \bf A}}+{\rm {\bf {{p}''}k\bm\sigma 
						\bf A\bm\sigma \bf k}}} \right)}{\left[ {\left( {2m-k^{0}} \right)k^{0}+{\rm {\bf 
						k}}^{2}} \right]^{2}}. \\ 
	\end{array}
\end{equation}
The equality of ${p}'=p^{IV} $ in Figs. 5(c) and 5(d) and equality of 
${p}'={{{p}'''}}$ in Figs. 5(g) and 5(h) do not lead to appearance of singular 
denominators in (\ref{eq51}), (\ref{eq52}). As a result, contributions of (\ref{eq51}), (\ref{eq52}) are completely compensated after renormalization of mass by appropriate mass counterterms.

\subsection{Contribution of Figs. 5(a) and 5(b)}

\begin{equation}
	\label{eq54}
	\begin{array}{l}
		\left( {\Lambda^{i}A^{i}} \right)_{5a+5b} =\left\{ {\left[ {V_{1}^{0\,} 
				\left( {{p}',{{{p}'''}}} \right)V_{1}^{0} \left( {{{{p}'''}},p^{IV}} 
				\right)V_{1}^{i} \left( {p^{IV},{{p}''}} \right)A^{i}} \right.} \right. \\ [12pt]
		\left. {-V_{1}^{l\,} \left( {{p}',{{{p}'''}}} \right)V_{1}^{l} \left( 
			{{{{p}'''}},p^{IV}} \right)A^{i}V_{1}^{i\,} \left( {p^{IV},{{p}''}} \right)} \right]  \\ [12pt]
	\left. { \times \dfrac{1}{\left( {{{{p}'''}}} \right)^{\,2}-m^{2}}+V_{2}^{\,00} \left( 
		{{p}',p^{IV}} \right)V_{1}^{i} \left( {p^{IV},{{p}''}} \right)A^{i} } \right.\\ [12pt]
		\left. {-V_{2}^{ll\,} \left( {{p}',p^{IV}} \right)V_{1}^{i} \left( 
			{p^{IV},{{p}''}} \right)A^{i}} \right\} \dfrac{1}{\left( {p^{IV}} 
			\right)^{2}-m^{2}} \\ [12pt]
		=e^3 \left\{ {\left[ {\left( {{p}'_{0} +{p}_{0}''' \,\,-\dfrac{{\rm {\bf 
								k}}^{2}}{III\left( {I+III} \right)}} \right) } \right. } \right. \\ [12pt]
	\left. { \left. { \times \left( {{p}_{0}''' 
					+p_{0}^{IV} \,\,\,-\dfrac{{\rm {\bf k}}^{2}}{III\left( {III+IV} \right)}} 
				\right)-\dfrac{3I\cdot IV}{III^{2}}{\rm {\bf k}}^{2}} \right]\,} \right. \\ [12pt]
			\times \left. \dfrac{1}{\left( {{{{p}'''}}} \right)^{\,2}-m^{2}}-1- {\dfrac{3I\cdot IV}{III^{2}}+\dfrac{{\rm {\bf k}}^{2}}{III^{2}\left( 
				{I+III} \right)\left( {III+IV} \right)}} \right\} \\ [12pt]
			\times \left( {-\dfrac{II}{IV}{\rm 
				{\bm \sigma \bf p}}^{IV}{\rm {\bm \sigma \bf A}}-\dfrac{IV}{II}{\rm {\bm \sigma 
					\bf A\bm\sigma \bf {{p}''}}}} \right)\dfrac{1}{\left( {p^{IV}} \right)^{2}-m^{2}}. \\ 
	\end{array}
\end{equation}
After algebraic transformations:
\begin{equation}
	\label{eq55}
	\begin{array}{l}
		\left( {\Lambda^{i}A^{i}} \right)_{5a+5b} =\left[ {{p}'_{0} {p}_{0}''' +{p}'_{0} p_{0}^{IV} }  +{p}_{0}''' +p_{0}^{IV}  \right. \\ [12pt]
		\left. { -\dfrac{{\rm {\bf k}}^{2}}{III\left( {I+III} \right)}\left( {{p}_{0}''' +p_{0}^{IV} } \right)- \dfrac{{\rm {\bf k}}^{2}}{III\left( {III+IV} 
			\right)}\left( {{p}'_{0} \,\,+{p}_{0}''' \,\,} \right)} \right.  \\ [12pt]
		\left. {+{\rm {\bf k}}^{2}+m^{2}-3I\cdot IV\left( {III^{2}-I^{2}} 
			\right)+\dfrac{{\rm {\bf k}}^{2}\left( {III-I} \right)}{III+IV}} 
		\right] \left( {-\dfrac{II}{IV}{\rm {\bm \sigma \bf p}}^{IV}{\rm {\bm \sigma 
					\bf A}}-\dfrac{IV}{II}{\rm {\bm \sigma \bf A\bm\sigma \bf {{p}''}}}} \right) \\ [12pt]
		\times \dfrac{1}{\left( {{{{p}'''}}} \right)^{\,2}-m^{2}}\dfrac{1}{\left( 
			{p^{IV}} \right)^{2}-m^{2}}=f\left( {{p}'_{\mu } -k_{\mu } ,{p}'_{\mu } } 
		\right)\dfrac{1}{\left( {p^{IV}} \right)^{2}-m^{2}}. \\ 
	\end{array}
\end{equation}
In (\ref{eq54}), (\ref{eq55}) $p^{IV}={p}'$, and we encounter the problem of singular 
denominator in expression $\frac{\mathcal{P}}{\left( {{p}'} 
	\right)^{\,2}-m^{2}}$, where $\mathcal{P}$ is the sign of principal value. To 
solve the problem, we use the Heitler limiting process \cite{bib8}.

Let us perform substitutions of
\begin{equation}
	\label{eq56}
	{{p}}_{\mu }''' \to {p}'_{\mu } \left( {1+{\varepsilon }'} \right)-k_{\mu 
	} ,
\end{equation}
\begin{equation}
	\label{eq57}
	p_{\mu }^{IV} \,\,\,\to {p}'_{\mu } \left( {1+\varepsilon } \right),
\end{equation}
where invariants ${\varepsilon }',\varepsilon $ tend to zero.

Now, let us write down Eq. (\ref{eq55}) as
\begin{equation}
	\label{eq58}
	\left( {\Lambda^{i}A^{i}} \right)_{5a+5b} =\int {f\left( {{p}'_{\mu } 
			\left( {1+{\varepsilon }'} \right)-k_{\mu } ,\,\,{p}'_{\mu } \left( 
			{1+\varepsilon } \right)} \right)\frac{\mathcal{P}}{m^{2}\left( {2\varepsilon 
				+\varepsilon^{2}} \right)}} \delta \left( {\varepsilon -{\varepsilon }'} 
	\right)\delta \left( {{\varepsilon }'} \right)d\varepsilon d{\varepsilon 
	}'.
\end{equation}
To eliminate divergent expressions, let us subtract the mass counterterm 
\begin{equation}
	\label{eq59}
	\left( {\Lambda^{i}A^{i}} \right)_{m} =\int {f\left( {{p}'_{\mu } -k_{\mu } 
			,\,\,{p}'_{\mu } \left( {1+\varepsilon } \right)} 
		\right)\frac{\mathcal{P}}{m^{2}\left( {2\varepsilon +\varepsilon^{2}} 
			\right)}} \delta \left( \varepsilon \right)d\varepsilon .
\end{equation}
In (\ref{eq58}) and (\ref{eq59})
\begin{equation}
	\label{eq60}
	\frac{\mathcal{P}}{2\varepsilon +\varepsilon 
		^{2}}=\frac{1}{2}\frac{\mathcal{P}}{\varepsilon 
	}-\frac{1}{2}\frac{1}{2+\varepsilon },
\end{equation}
\begin{equation}
	\label{eq61}
	\frac{1}{2}\frac{\mathcal{P}}{\varepsilon }\delta \left( \varepsilon 
	\right)=-\,\,\frac{1}{4}{\delta }'\left( \varepsilon \right).
\end{equation}
In (\ref{eq55}) and (\ref{eq59})
\begin{equation}
	\label{eq62}
	\left. {f\left( {{p}'_{\mu } \left( {1+{\varepsilon }'} \right)-k_{\mu } 
			,\,\,{p}'_{\mu } \left( {1+\varepsilon } \right)} \right)} 
	\right|_{{\varepsilon }',\,\varepsilon =0} 
	=- e^3 \frac{4m^{2}+4mk^{0}}{\left( {-\left( {2m-k^{0}} \right)k^{0}-{\rm 
				{\bf k}}^{{\rm {\bf 2}}}} \right)}L.
\end{equation}
In what follows, let us denote $M^{2}=4m^{2}+4mk^{0}$.

Upon integration in (\ref{eq58}) on ${\varepsilon }'$, the total contribution of 
(\ref{eq58}) and (\ref{eq59}), taking into account Eqs. (\ref{eq60}) and (\ref{eq61}), can be written as
\begin{equation}
	\label{eq63}
	\begin{array}{l}
		\left( {\Lambda^{i}A^{i}} \right)_{5a+5b+m} =-\dfrac{1}{4m^{2}}\int 
		{f\left( {{p}'_{\mu } \left( {1+\varepsilon } \right),\,{p}'_{\mu } } 
			\right)\times } \,{\delta }'\left( \varepsilon \right)d\varepsilon  \\ [12pt]
		=\left. {\dfrac{1}{4m^{2}}\dfrac{\partial }{\partial \varepsilon }f\left( 
			{{p}'_{\mu } \left( {1+\varepsilon } \right),\,{p}'_{\mu } } \right)} 
		\right|_{\varepsilon =0} \\ 
	\end{array}
\end{equation}
Likewise, we can calculate the contribution of diagrams 5(e) and 5(f). The total 
contribution of these diagrams together with the contribution of diagrams 
(\ref{eq63}) is
\begin{equation}
	\label{eq64}
	\left( {\Lambda^{i}A^{i}} \right)_{SE} =2 e^3 \frac{2m^{2}-2mk^{0}-\left( 
		{k^{0}} \right)^{2}-{\rm {\bf k}}^{{\rm {\bf 2}}}}{\left( {-2mk^{0}+\left( 
			{k^{0}} \right)^{2}-{\rm {\bf k}}^{{\rm {\bf 2}}}} \right)^{2}}L.
\end{equation}
The total contribution of all renormalized Figs. 5(a)-5(l) is equal to the 
sum of expressions (\ref{eq50}) and (\ref{eq64})

\begin{equation}
	\label{eq65}
	\left( {\Lambda^{i}A^{i}} \right)_{amm} =e^3 \frac{\left( 
		{4m^{2}-4mk^{0}-2\left( {k^{0}} \right)^{2}-2{\rm {\bf k}}^{{\rm {\bf 2}}}} 
		\right)L-4{\rm {\bf kA}}\left( {{\rm {\bm \sigma \bf {p}'\bm\sigma \bf k}}+{\rm {\bm 
					\sigma \bf k\bm\sigma \bf {{p}''}}}} \right)}{\left( {-2mk^{0}+\left( {k^{0}} 
			\right)^{2}-{\rm {\bf k}}^{{\rm {\bf 2}}}} \right)^{2}}.
\end{equation}
The expression (\ref{eq65}) coincides with the similar expression in Dirac 
representation \cite{bib8}. Expression (\ref{eq65}) allows to 
calculate of the anomalous magnetic moment of the electron in the low order 
of the perturbation theory. The final results agree with the computational 
results for the anomalous magnetic moment of the electron in the Dirac 
representation.

\section{Lamb shift of energy levels of atomic electrons}

The Feynman diagrams needed to calculate the Lamb shift are presented in 
Fig. 5.

In the process under consideration, static electromagnetic potentials 
are \\ $A^{i}\left( {{\rm {\bf x}}} \right)=0$. $A^{0}\left( {{\rm {\bf x}}} 
\right)={Ze} \mathord{\left/ {\vphantom {{Ze} {4\pi \left| {{\rm {\bf x}}} 
				\right|}}} \right. \kern-\nulldelimiterspace} {4\pi \left| {{\rm {\bf x}}} 
	\right|}$ is the Coulomb potential of the nucleus with atomic number $Z$.

The amplitude of the process in the first order of the perturbation theory 
is (see in Fig. 1 and \eqref{appB.1})
\begin{equation}
	\label{eq66}
	S_{fi} =-e\frac{i\delta \left( {E_{f} -E_{i} } \right)}{\left( {2\pi } 
		\right)^{2}2E_{i} }A^{0}\left( {{\rm {\bf q}}} \right)\bar{{U}}_{S_{f} } 
	\left\langle {{\rm {\bf p}}_{{\rm {\bf f}}} \left| {E_{i} +m+\frac{1}{E_{i} 
				+m}{\rm {\bm \sigma \bf p}}_{{\rm {\bf f}}} {\rm {\bm \sigma \bf p}}_{{\rm {\bf i}}} 
		} \right|{\rm {\bf p}}_{{\rm {\bf i}}} } \right\rangle U_{S_{i} } .
\end{equation}
Here,  $A^{0}\left( {{\rm {\bf q}}} \right)={4\pi } \mathord{\left/ 
	{\vphantom {{4\pi } {\left( {{\rm {\bf q}}^{{\rm {\bf 2}}}} \right)}}} 
	\right. \kern-\nulldelimiterspace} {\left( {{\rm {\bf q}}^{{\rm {\bf 2}}}} 
	\right)}$.

In our denotations, 
\begin{equation*}
	\begin{array}{l}
	{\rm {\bf p}}_{{\rm {\bf f}}} ={\rm {\bf 
		{p}'}},\,\,{\rm {\bf p}}_{{\rm {\bf i}}} ={\rm {\bf {{p}''}}},\,\,{\rm {\bf 
		{{{p}'''}}}}={\rm {\bf {p}'}}-{\rm {\bf k}},\,\,{\rm {\bf p}}^{{\rm {\bf 
			IV}}}={\rm {\bf {{p}''}}}-{\rm {\bf k}}, \\ [12pt]
		q={p}'-{{p}''}=\left\{ 
{\begin{array}{l}
		q^{0}=0,\,\,\left( {{p}'} \right)^{0}=\left( {{{p}''}} \right)^{0}, \\ 
		q^{i}=\,\left( {{p}'} \right)^{i}-\left( {{{p}''}} \right)^{i}. \\ 
\end{array}} \right.
\end{array}
	\end{equation*}

In the following, we will consider electron motion in the Coulomb field in the 
non-relativistic approximation and take into account the addends up to 
$\left( {{\rm {\bf {p}'}}} \right)^{2},\,\,\left( {{\rm {\bf {{p}''}}}} 
\right)^{2}$.

Let us consider the contribution of Figs. 5(i)--5(l) to the Lamb 
shift (see Fig. 5). For this type of diagrams the equality of $III=IV_{\, 
}$ holds true. We write down the amplitude as
\begin{equation}
	\label{eq67}
	-e\frac{i\delta \left( {{E}'-{{E}''}} \right)}{\left( {2\pi } 
		\right)^{2}2{{E}''}}\bar{{U}}_{{S}'} \left\langle {{\rm {\bf {p}'}}\left| 
		{\frac{1}{\left( {2\pi } \right)^{4}}\int {\frac{d^{4}k}{k^{2}}\Lambda 
				^{0}\left( {{\rm {\bf {p}'}},{\rm {\bf {{p}''}}}} \right)A^{0}\left( {{\rm 
						{\bf q}}} \right)} } \right|{\rm {\bf {{p}''}}}} \right\rangle U_{S'' } .
\end{equation}

\subsection{Contribution of Fig. 5(i)}

\begin{equation}
	\label{eq68}
	\begin{array}{l}
		\left( {\Lambda^{0}A^{0}} \right)_{5i} =\left[ {V_{1}^{0} \left( 
			{{p}',{{{p}'''}}} \right)} \right.V_{1}^{0} \left( {{{{p}'''}},p^{IV}} 
		\right)A^{0}V_{1}^{0} \left( {p^{IV},{{p}''}} \right) \\ [12pt]
		\left. {-V_{1}^{i} \left( {{p}',{{{p}'''}}} \right)V_{1}^{0} \left( 
			{{{{p}'''}},p^{IV}} \right)A^{0}V_{1}^{i} \left( {p^{IV},{{p}''}} \right)} 
		\right]\dfrac{1}{\left( {\left( {{{{p}'''}}} \right)^{2}-m^{2}} \right)\left( 
			{\left( {p^{IV}} \right)^{2}-m^{2}} \right)}. \\ 
	\end{array}
\end{equation}
The expression (\ref{eq68}) with substitution of $V_{1}^{0} ,\,\,V_{1}^{i} $ has a cumbersome form and is given in the App. C (see \eqref{appC.1}).

\subsection{Contribution of Figs. 5(j) and 5(k)}

\begin{equation}
	\label{eq69}
	\begin{array}{l}
		\left( {\Lambda^{0}A^{0}} \right)_{5j} +\left( {\Lambda^{0}A^{0}} 
		\right)_{5k} =\left[ {V_{1}^{0} \left( {{p}',{{{p}'''}}} \right)} 
		\right.V_{2}^{0\,0} \left( {{{{p}'''}},{{p}'}'} \right)A^{0} \\ [12pt]
		\left. {-V_{1}^{i} \left( {{p}',{{{p}'''}}} \right)V_{2}^{0i} \left( 
			{{{{p}''}},{{p}''}} \right)A^{0}} \right]\dfrac{1}{\left( {{{{p}'''}}} 
			\right)^{2}-m^{2}}+\left[ {V_{2}^{00} \left( {{p}',p^{IV}} \right)} 
		\right.A^{0}V_{1}^{\,0} \left( {p^{IV},{{p}''}} \right) \\ [12pt]
		-\left. {V_{2}^{i0} \left( {{p}',p^{IV}} \right)A^{0}V_{1}^{i} \left( 
			{p^{IV},{{p}''}} \right)} \right]\dfrac{1}{\left( {p^{IV}} 
			\right)^{2}-m^{2}}. \\ 
	\end{array}
\end{equation}

\subsection{Contribution of Fig. 5(l)}

\begin{equation}
	\label{eq70}
	\left( {\Lambda^{0}A^{0}} \right)_{5l} =V_{3}^{000} A^{0}-V_{3}^{i0i} 
	A^{0}.
\end{equation}
The final expressions (\ref{eq69}), (\ref{eq70}) are given in the App. C (see Eqs. \eqref{appC.2}, \eqref{appC.3}).

The sum of \eqref{appC.1}--\eqref{appC.3} accurate to $\left( {{\rm {\bf {p}'}}} 
\right)^{2},\left( {{\rm {\bf {{p}''}}}} \right)^{2}$ was determined by 
using software package ''Maple''. As a result, the total contribution of 
Figs. 5(i)--5(l) can be written as

\begin{equation}
	\label{eq71}
	\begin{array}{l}
		\left( {\Lambda^{0}A^{0}} \right)_{5i+5j+5k+5l} =2 e^3 m\left[ {4m^{2}-2\left( 
			{k^{0}} \right)^{2}-2{\rm {\bf k}}^{2}-4mk^{0}} \right. \\ [12pt]
		-\dfrac{k^{2}+\left( {k^{0}} \right)^{2}+2mk^{0}-10m^{2}}{4m^{2}}\left( 
		{\left( {{\rm {\bf {p}'}}} \right)^{2}+\left( {{\rm {\bf {{p}''}}}} 
			\right)^{2}} \right)-4{\rm {\bf {p}'{{p}''}}}+4\left( {{\rm {\bf 
					{p}'k}}+{\rm {\bf {{p}''}k}}} \right) \\ [12pt]
		\left. {+\left( {\dfrac{2k^{0}}{m}-4} \right)\left( {{\rm {\bm \sigma 
					\bf {p}'\bm\sigma \bf k}}+{\rm {\bm \sigma \bf k \bm\sigma \bf {{p}''}}}} 
			\right) } \right. \\ [12pt]
			\left. { + \dfrac{1}{4m^{2}}\left( {4m^{2}-2\left( {k^{0}} \right)^{2}-2{\rm 
					{\bf k}}^{2}-4mk^{0}} \right){\rm {\bm \sigma \bf {p}'\bm\sigma \bf {{p}''}}}} \right]A^0 \left( {\bf q} \right) . 
		\\ 
	\end{array}
\end{equation}
In the standard QED, the similar contribution to the Lamb shift is provided 
by the only 5(i) with vertexes $-ie\gamma^{\mu }$. The analog of (\ref{eq71}) 
is (see, for example, Ref. \cite{bib12})
\begin{equation}
	\label{eq72}
	\begin{array}{l}
	\left( {\Lambda^{0}A^{0}} \right)_{Dir}  =e^3 \left\lbrace \left( {4m^{2}+2{\rm {\bf 
				q}}^{{\rm {\bf 2}}}-2{\rm {\bf k}}^{{\rm {\bf 2}}}-2\left( {k^{0}} 
		\right)^{2}} \right)\gamma^{0} \right. \\ [12pt]
  \left. 	+ 4i{\rm {\bm \sigma }}\left( {{\rm {\bf 
				k}}\times {\rm {\bf q}}} \right)\gamma^{0}+k^{0}\left( {-4m+4{\rm {\bm 
				\gamma\bf  k}}} \right) \right\rbrace A^0 \left( \bf q\right) .
				\end{array}
\end{equation}
In the scattering amplitude, expression (\ref{eq72}) will be found sandwiched between bispinors 
$\Psi \left( {{{p}''}} \right)$ and $\bar{{\Psi }}\left( {{p}'} 
\right)\gamma^{0}$, where
\begin{equation}
	\label{eq73}
	\Psi \left( {{{p}''}} \right)=\frac{1}{\sqrt {2\left( {{{p}''}} \right)^{0}} 
	}\sqrt {\left( {{{p}''}} \right)^{0}+m} \left( {{\begin{array}{*{20}c}
				\,\,\,\,\,\,\,\,\,\,\,\,\,\,{\varphi \left( {{{p}''}} \right)} \hfill \\ [5pt]
				{\dfrac{{\rm {\bm \sigma \bf {{p}''}}}}{\left( {{{p}''}} \right)^{0}+m}\varphi 
					\left( {{{p}''}} \right)} \hfill \\
	\end{array} }} \right).
\end{equation}
Expanding (\ref{eq73}) and $\bar{{\Psi }}\left( {{p}'} \right)\gamma^{0}$ in powers not higher than $\left( {{\rm {\bf {p}'}}} \right)^{2},\left( {{\rm 
		{\bf {{p}''}}}} \right)^{2}$ and considering, with our accuracy, 
that spinors are  $\varphi \left( {{p}'} \right)=2m\,\Phi \left( {{p}'} 
\right),\,\,\varphi \left( {{{p}''}} \right)=2m\,\Phi \left( {{{p}''}} 
\right)$ (see (\ref{eq10}) and (\ref{eq11})), we obtain that the expression (\ref{eq72}) sandwiched between spinors $\Phi \left( {{p}'} \right),\,\,\Phi \left( {{{p}''}} \right)$ coincides with (\ref{eq71}).

Let us note that expression $\dfrac{1}{\sqrt {\left( 
		{{{p}''}} \right)^{0}} }$ in (\ref{eq73}), as well as expression $\dfrac{1}{\sqrt 
	{\left( {{p}'} \right)^{0}} }$ in Hermitian-conjugated spinor are not subjected to expansion since in amplitude (\ref{eq67}) both the expressions are present in multiplied form 
of $\dfrac{1}{2{{E}''}}$.

\subsection{Contribution of Figs. 5(a)--5(h)}

Now, let us consider to the contribution of self-energy diagrams 5(a)--5(h). 
For Figs. 5(a)--5(d) equalities $I=II=IV_{\, }$ hold true. For Figs. 5(e)--5(h), equalities $I=II=III$ hold true. The equality ${p}'=p^{IV}$ in 
diagrams 5(c) and 5(d) and equality ${{p}''}={{{p}'''}}$ in Figs. 5(g) and 5(h) do not 
lead to appearance of singular denominators in calculations. As a result, 
the contributions of Figs. 5(c), 5(d), and  5(g), 5(h) are completely compensated for 
after renormalization of mass by appropriate mass counterterms. The 
contribution of Figs. 5(a), 5(b), and 5(e), 5(f) is determined by the expression
\begin{equation}
	\label{eq74}
	\begin{array}{l}
		\left( {\Lambda^{0}A^{0}} \right)_{5a+5b+5e+5f} =\left\{ {\left[ 
			{V_{1}^{0} \left( {{p}',{{{p}'''}}} \right)V_{1}^{0} \left( 
				{{{{p}'''}},p^{IV}} \right)-V_{1}^{i} \left( {{p}',{{{p}'''}}} 
				\right)V_{1}^{i} \left( {{{{p}'''}},p^{IV}} \right)} \right] } \right. 
		\\ [12pt]
		\times \dfrac{1}{\left( {{{{p}'''}}} \right)^{2}-m^{2}}+V_{2}^{00} \left( 
		{{p}',p^{IV}} \right)-\left. {V_{2}^{ll} \left( {{p}',p^{IV}} \right)} 
		\right\} \dfrac{V_{1}^{0} \left( {p^{IV},{{p}''}} \right)A^{0}}{\left( 
			{p^{IV}} \right)^{2}-m^{2}} \\ [12pt]
		+\left\{ {\left[ {V_{1}^{0} \left( {{{{p}'''}},p^{IV}} \right)V_{1}^{0} 
				\left( {p^{IV},{{p}''}} \right)-V_{1}^{i} \left( {{{{p}'''}},p^{IV}} 
				\right)V_{1}^{i} \left( {p^{IV},{{p}''}} \right)} \right]}  \right. \\ [12pt]
	\left. { \times \dfrac{1}{\left( {p^{IV}} \right)^{2}-m^{2}}+V_{2}^{00} \left( {{{{p}'''}},{{p}''}} \right)-V_{2}^{ll} \left( 
			{{{{p}'''}},{{p}''}} \right)} \right\} \dfrac{V_{1}^{0} \left( 
			{{p}',{{{p}'''}}} \right)A^{0}}{\left( {{{{p}'''}}} \right)^{2}-m^{2}}. \\ 
	\end{array}
\end{equation}
For Figs. 5(a), 5(b), and 5(e), 5(f) , there is a problem of singular denominators. 
In expression (\ref{eq74}), these denominators stand after curly brackets. To 
solve the problem, we used the Heitler limiting process \cite{bib8} as in Sec. 5. Performing calculations similar 
to the calculations in Sec. 5 (see (\ref{eq54})--(\ref{eq64})), we will obtain the total 
contribution of self-energy diagrams
\begin{equation}
	\label{eq75}
	\left( {\Lambda^{0}A^{0}} \right)_{SE} = 2e^3\frac{2m^{2}+{\rm {\bf k}}^{{\rm 
				{\bf 2}}}-2\left( {{p}'} \right)^{0}k^{0}+2{\rm {\bf 
				{p}'k}}-\dfrac{2}{m^{2}}\left( {\left( {{p}'} \right)^{0}k^{0}-{\rm {\bf 
					{p}'k}}} \right)^{2}}{\left[ {\left( {\left( {{p}'} \right)^{0}-k^{0}} 
			\right)^{2}-\left( {{\rm {\bf {p}'}}-{\rm {\bf k}}} \right)^{2}-m^{2}} 
		\right]^{2}} A^0 \left( \bf q\right) .
\end{equation}
The expression (\ref{eq75}) coincides with the similar expression in Dirac 
representation (see for example, Ref. \cite{bib8}). The total 
contribution of Figs. 5(a)--5(h) to the Lamb shift of atomic levels 
coincides with the similar contribution of diagrams in the Dirac 
representation with vertexes $-ie\gamma^{\mu }$.

\subsection{Contribution of Figs. 5(m)--5(o)}

Diagrams 5(m)--5(o) are connected with self-energy function of photon 
(see, for example, Ref. \cite{bib16}). In the second order of the 
perturbation theory, the photon propagator is written in the form of
\begin{equation}
	\label{eq76}
	-iD_{\alpha \beta } =\frac{\left( {-i} \right)g_{\alpha \beta } 
	}{q^{2}}+\frac{\left( {-i} \right)g_{\alpha \mu } }{q^{2}}i\Pi_{\mu \nu } 
	\frac{\left( {-i} \right)g_{\nu \beta } }{q^{2}}.
\end{equation}
In compliance with diagrams 5(m)--5(o), the tensor $\Pi_{\mu \nu } $ has the 
form of
\begin{equation}
	\label{eq77}
	\begin{array}{l}
		\Pi_{\mu \nu } \left( q \right)=i4\pi e^{2}\int {\dfrac{d^{4}p}{\left( 
				{2\pi } \right)^{4}}Tr\left[ {\dfrac{V_{1\mu } \left( {p,p-q} \right)V_{1\nu 
					} \left( {p-q,p} \right)+V_{1\mu } \left( {p-q,p} \right)V_{1\nu } \left( 
					{p,p-q} \right)}{2\left( {p^{2}-m^{2}} \right)\left( {\left( {p-q} 
						\right)^{2}-m^{2}} \right)}} \right.}  \\ [12pt]
		\left. {+\dfrac{V_{2\mu \nu } \left( {p-q,p,p-q} \right)}{\left( {p-q} 
				\right)^{2}-m^{2}}+\dfrac{V_{2\mu \nu } \left( {p,p-q,p} 
				\right)}{p^{2}-m^{2}}} \right]. \\ 
	\end{array}
\end{equation}
At electron scattering in the static external field $A_{0} \left( {{\rm {\bf 
			x}}} \right)$ only the component $\Pi_{00} \left( q \right)$ contributes to 
amplitude of process. In this case $q_{0} =0$. Then 
\begin{equation}
	\label{eq78}
	\begin{array}{l}
		f\left( {p^{0}} \right)=Tr\left[ {V_{10} \left( {p,p-q} \right)V_{10} 
			\left( {p-q,p} \right)} \right]=Sp\left[ {V_{10} \left( {p-q,p} 
			\right)V_{10} \left( {p,p-q} \right)} \right] \\ [12pt]
		= 2 e^2 \left[ {\left( {2p^{0}-\dfrac{{\rm {\bf p}}^{2}}{2\left( {p^{0}+m} 
					\right)}-\dfrac{{\rm {\bf p}}_{1}^{2} }{2\left( {p^{0}+m} \right)}} 
			\right)^{2}}\right.\\ [12pt]
		\left. {+ \dfrac{2{\rm {\bf pp}}_{1} }{p^{0}+m}\left( {2p^{0}-\dfrac{{\rm 
						{\bf p}}_{1}^{2} }{2\left( {p^{0}+m} \right)}-\dfrac{{\rm {\bf 
							p}}^{2}}{2\left( {p^{0}+m} \right)}} \right)+\dfrac{{\rm {\bf p}}^{2}{\rm 
					{\bf p}}_{1}^{2} }{\left( {p^{0}+m} \right)^{2}}} \right], \\ 
	\end{array}
\end{equation}
\begin{equation}
	\label{eq79}
	TrV_{2}^{00} \left( {p,p-q,p} \right)= 2 e^2 \left[ {-1+\frac{3}{4}\frac{{\rm {\bf 
					p}}^{2}}{\left( {p^{0}+m} \right)^{2}}+\frac{1}{4}\frac{{\rm {\bf 
					p}}_{1}^{2} }{\left( {p^{0}+m} \right)^{2}}-\frac{{\rm {\bf pp}}_{1} 
		}{\left( {p^{0}+m} \right)^{2}}} \right],
\end{equation}
\begin{equation}
	\label{eq80}
	TrV_{2}^{00} \left( {p-q,p,p-q} \right)= 2 e^2 \left[ {-1+\frac{1}{4}\frac{{\rm 
				{\bf p}}^{2}}{\left( {p^{0}+m} \right)^{2}}+\frac{3}{4}\frac{{\rm {\bf 
					p}}_{1}^{2} }{\left( {p^{0}+m} \right)^{2}}-\frac{{\rm {\bf pp}}_{1} 
		}{\left( {p^{0}+m} \right)^{2}}} \right].
\end{equation}
In (\ref{eq78}) - (\ref{eq80})
\begin{equation}
	\label{eq81}
	\begin{array}{l}
		\Pi_{00} \left( q \right)=-\dfrac{4\pi e^{2}}{\left( {2\pi } 
			\right)^{3}}\int {d{\rm {\bf p}}\,} Tr\left\{ {\dfrac{f\left( {p^{0}=E_{1} } 
				\right)}{2E_{1} \left( {E_{1}^{2} -E^{2}} \right)}-} \right.\dfrac{f\left( 
			{p^{0}=E} \right)}{2E\left( {E_{1}^{2} -E^{2}} \right)} \\ [12pt]
		+\left. {\left. {V_{2}^{00} \left( {p-q,p,p-q} \right)} 
			\right|_{p^{0}=E_{1} } \dfrac{1}{2E_{1} }+\left. {V_{2}^{00} \left( {p,p-q,p} 
				\right)} \right|_{p^{0}=E} \dfrac{1}{2E}} \right\} . \\ 
	\end{array}
\end{equation}
The expression (\ref{eq81}) is obtained by contour integration of (\ref{eq77}) with respect 
to variable d$p^{0}$ with bypass of poles with positive energy 
$p^{0}=E=\left( {m^{2}+{\rm {\bf p}}^{2}} \right)^{1/2}$ and $p^{0}=E_{1} 
=\left( {m^{2}+\left( {{\rm {\bf p}}-{\rm {\bf q}}} \right)^{2}} \right)^{1/2}$.

\mbox{After algebraic transformations, the expression (\ref{eq81}) can be written in the form of} 
\begin{equation}
	\label{eq82}
	\Pi_{00} \left( q \right)=\frac{e^{2}}{\pi^{2}}\int {d{\rm {\bf 
				p}}\frac{-EE_{1} +{\rm {\bf pp}}_{1} +m^{2}}{EE_{1} \left( {E+E_{1} } 
			\right)}} \,.
\end{equation}
The expression (\ref{eq82}) coincides with the tensor $\Pi_{00} \left( q \right)$ 
in standard QED (see, for example, Ref. \cite{bib8}). 

Thus, the total contribution of diagrams 5(a)--5(o) in the Lamb shift of 
atomic levels coincides with the appropriate contribution of diagrams in the 
standard QED with vertices $-ie\gamma^{\mu }.$

In the developed theory, it is not necessary to include creation and annihilation of virtual electron-positron pairs in the calculations while defining physical effects. Let us adduce some arguments.

\begin{enumerate}
	\item The final calculation results of Figs. 5(m)--5(o) connected with self-energy function of a photon are obtained by using only states with positive energies of electrons in intermediate states.
	\item Diagram 5(m) can be considered as vacuum polarization by virtual electron-positron pairs but the Figs. 5(n) and 5(o) cannot be interpreted in such manner. However these two diagrams are required to determine, for example, Lamb shift at using Klein-Gordon equation with spinor functions for fermions \cite{bib9,bib17}.
\end{enumerate}

\section{Conclusions}
In the paper, quantum electrodynamics is considered with 
self-conjugated equations with spinor wave functions for fermion fields.

For the considered QED, some of Feynman diagrams are calculated. In the 
low order of the perturbation theory, the matrix elements of Coulomb 
scattering of electrons, scattering of an electron on a proton (M{\o}ller 
scattering), Compton effect, and annihilation of the electron-positron pair 
are calculated. The final results coincide with the similar values 
calculated in the standard QED when using the Dirac equation with the 
bispinor wave functions.

Self-energy of an electron with ultraviolet logarithmic divergence coincides 
with the value computed in the standard QED. New feature is the absence of 
contribution to the self-energy of intermediate states with negative energies. 
In the standard QED in this case, the self-energy linearly diverges in the 
ultraviolet limit. Only accounting of the contribution of the intermediate 
states with positive and negative energies leads to the ultraviolet logarithmic divergence 
of self-energy \cite{bib8}. The only intermediate states 
with positive energies are also used while calculating the matrix elements 
related to the anomalous magnetic moment and the Lamb shift of electron.

The calculated values of the matrix elements for determining the anomalous 
magnetic moment of the electron and Lamb shift are consistent with the 
matrix elements determined in the standard QED.

In the QED under consideration, there are no operators connecting solutions 
with positive and negative energies for fermions. The equations for 
electrons and positrons are not connected with each other.

In the developed theory, while defining the physical effects, we do not need to introduce the concept of vacuum polarization and calculate creation and annihilation of virtual electron-positron pairs. In QED under consideration, in the equations, the masses of particles and antiparticles have the opposite signs. For the first time one of the authors showed this in 1989 (see Ref. \cite{bib13}--\cite{bib15}). Later, other researchers came to the same conclusion (see, for example, Refs. \cite{bib18}--\cite{bib20}). In a certain sense, emergence of negative mass in our theory is the result of refusal to use states with negative energies (see Eqs. (12), (14) or (13), (15)). The future will show possibility of proving a more profound physical significance of this fact. 

\section*{Acknowledgments}

The authors express their gratitude to A.L.Novoselova for essential 
technical assistance in preparation of the paper.

\section*{APPENDIX A. Interaction operators $\sim e^{2},\,\,e^{3}$}

\[	
\begin{array}{l} 
	\label{appA.1}
	V_{2} =e^{2}\int {d{\rm {\bf {{{p}'''}}}}\left\{ {\left[ 
			{-1-\dfrac{\left( {{\rm {\bf {p}'}}} \right)^{2}}{I\left( {I+II} 
					\right)\left( {I+III} \right)\left( {III+II} \right)}} \right. } \right.} \\ [12pt]
		\left. {+\dfrac{\left( {{\rm {\bf {{{p}'''}}}}} \right)^{2}}{III^{2}\left( {I+III} \right)\left( 
				{II+III} \right)}} -\dfrac{\left( {{\rm {\bf {{p}''}}}} \right)^{2}}{II\left( {I+II} 
		\right)\left( {I+III} \right)\left( {III+II} \right)} \right.\\ [12pt]
	\left. {+\dfrac{{\rm {\bm \sigma \bf {p}'\bm\sigma \bf {{p}''}}}}{I\cdot II\cdot III^{2}}-\dfrac{{\rm {\bm \sigma \bf {p}'\bm\sigma \bf {{{p}'''}}}}}{II\cdot III^{2}\cdot \left( {II+III} 
			\right)}-\dfrac{{\rm {\bm \sigma \bf {{{p}'''}}\bm\sigma \bf {{p}''}}}}{II\cdot 
			III^{2}\cdot \left( {I+III} \right)}} \right] \\  [15pt]
	\times \left\langle {{\rm {\bf {p}'}}\left| {A^{0}} \right|{\rm {\bf 
				{{{p}'''}}}}} \right\rangle \left\langle {{\rm {\bf {{{p}'''}}}}\left| 
		{A^{0}} \right|{\rm {\bf {{p}''}}}} \right\rangle +\dfrac{I\cdot 
		II}{III^{2}}\sigma^{i}\sigma^{k}\left\langle {{\rm {\bf {p}'}}\left| 
		{A^{i}} \right|{\rm {\bf {{{p}'''}}}}} \right\rangle \left\langle {{\rm {\bf 
				{{{p}'''}}}}\left| {A^{k}} \right|{\rm {\bf {{p}''}}}} \right\rangle  \\  [15pt]
	+\left[ {\dfrac{{\rm {\bm \sigma \bf {p}'}} \sigma^{i}}{I\left( {I+III} 
			\right)}-\dfrac{I\sigma^{i}{\rm {\bm \sigma \bf {{p}''}}}}{II\cdot 
			III^{2}}+\dfrac{I\sigma^{i}{\rm {\bm \sigma \bf {{{p}'''}}}}}{III^{2}\left( 
			{III+II} \right)}} \right]\left\langle {{\rm {\bf {p}'}}\left| {A^{i}} 
		\right|{\rm {\bf {{{p}'''}}}}} \right\rangle \left\langle {{\rm {\bf 
				{{{p}'''}}}}\left| {A^{0}} \right|{\rm {\bf {{p}''}}}}  \right\rangle  \\  [15pt]
	+\left[ {-\,\dfrac{II}{I\cdot III^{2}}{\rm {\bm \sigma \bf {p}'}}\sigma 
		^{i}+\dfrac{1}{II\left( {I+III} \right)}\sigma^{i}{\rm {\bm \sigma 
				\bf {{p}''}}}+\dfrac{II}{III^{2}\left( {I+III} \right)}{\rm {\bm \sigma 
				\bf {{{p}'''}}}}\sigma^{i}} \right] \\  [15pt]
	\left. {\times \left\langle {{\rm {\bf {p}'}}\left| {A^{0}} \right|{\rm 
				{\bf {{{p}'''}}}}} \right\rangle \left\langle {{\rm {\bf {{{p}'''}}}}\left| 
			{A^{i}} \right|{\rm {\bf {{p}''}}}} \right\rangle } \right\} , \\ \tag{A.1}
\end{array}  
\]

\[ 
\begin{array}{l}
	V_{3} =\pm e^{3}\int d {\rm {\bf {{{p}'''}}}}d{\rm {\bf p}}^{IV}\left\{ 
	{\left( {\dfrac{\left( {{\rm {\bf p}}^{IV}} \right)^{2}}{IV^{2}\cdot \left( 
				{I+IV} \right)\left( {I+III} \right)\left( {III+IV} \right)\left( {IV+II} 
				\right)}} \right. }\right. \\ [15pt]
	\left.{	-\dfrac{{\rm {\bm \sigma \bf {p}'\bm\sigma \bf p}}^{IV}}{I\cdot 
			III^{2}\cdot IV^{2}\left( {IV+II} \right)} -\dfrac{{\rm {\bm \sigma \bf {{{p}'''}}\bm\sigma \bf {{p}''}}}}{II\cdot III^{2}\cdot 
			IV^{2}\left( {I+III} \right)} } \right. \\ [15pt]
	\left.{ +\dfrac{\left( {{\rm {\bf {{{p}'''}}}}} 
			\right)^{2}}{III^{2}\cdot \left( {I+III} \right)\left( {III+IV} 
			\right)\left( {III+II} \right)\left( {IV+II} \right)} }\right. \\ [15pt]
	-\dfrac{{\rm {\bm \sigma \bf p}}^{IV}{\rm {\bm \sigma \bf {{p}''}}}}{II\cdot 
		IV^{2}\left( {I+III} \right)\left( {I+IV} \right)\left( {III+IV} \right)}  \\[15pt]
	- \dfrac{{\rm {\bm \sigma \bf {p}'\bm\sigma \bf {{{p}'''}}}}}{I\cdot 
		III^{2}\left( {III+II} \right)\left( {III+IV} \right)\left( {IV+II} 
		\right) }\\ [12pt]
		+\dfrac{{\rm {\bm \sigma \bf {{{p}'''}}\bm\sigma p}}^{IV}}{III^{2}\cdot 
		IV^{2}\left( {I+III} \right)\left( {IV+II} \right)}+\dfrac{{\rm {\bm \sigma 
				\bf {p}'\bm\sigma {{p}''}}}}{I\cdot II\cdot III^{2}\cdot IV^{2}} \\ [12pt]
	-\dfrac{\left( {{\rm {\bf {p}'}}} \right)^{2}\left( {I+II+III+IV} 
		\right)}{I\left( {I+II} \right)\left( {I+III} \right)\left( {I+IV} 
		\right)\left( {III+IV} \right)\left( {III+II} \right)\left( {IV+II} 
		\right)} \\ [12pt]
	-\left. {\dfrac{\left( {{\rm {\bf {{p}''}}}} \right)^{2}\left( {I+II+III+IV} 
			\right)}{II\left( {I+II} \right)\left( {I+III} \right)\left( {I+IV} 
			\right)\left( {III+IV} \right)\left( {III+II} \right)\left( {IV+II} 
			\right)}} \right) \\ [12pt]
			\times \left\langle {{\rm {\bf {p}'}}\left| {A^{0}} \right|{\rm {\bf 
					{{{p}'''}}}}} \right\rangle \left\langle {{\rm {\bf {{{p}'''}}}}\left| 
			{A^{0}} \right|{\rm {\bf p}}^{IV}} \right\rangle \left\langle {{\rm {\bf 
					p}}^{IV}\left| {A^{0}} \right|{\rm {\bf {{p}''}}}} \right\rangle \\ [12pt]
				+ \dfrac{I\cdot II}{III^{2}\cdot IV^{2}}\sigma^{i}\sigma^{k}\left\langle 
				{{\rm {\bf {p}'}}\left| {A^{i}} \right|{\rm {\bf {{{p}'''}}}}} \right\rangle 
				\left\langle {{\rm {\bf {{{p}'''}}}}\left| {A^{0}} \right|{\rm {\bf 
							p}}^{IV}} \right\rangle \left\langle {{\rm {\bf p}}^{IV}\left| {A^{k}} \right|{\rm {\bf 
							{{p}''}}}} \right\rangle 
\end{array}
\]
\[
\begin{array}{l}
	\label{appA.2}
		+\left( {-\dfrac{{\rm {\bm \sigma \bf {{{p}'''}}}}\sigma 
			^{i}}{III^{2}\left( {I+III} \right)\left( {IV+II} \right)}}-\dfrac{\sigma^{i}{\rm {\bm \sigma \bf p}}^{IV}}{IV^{2}\left( {I+III} 
		\right)\left( {IV+II} \right) }\right. \\ [12pt]
	\left.{	+\dfrac{\sigma^{i}{\rm {\bm \sigma \bf {{p}''}}}}{II\cdot IV^{2}\left( {I+III} \right)}}  + \dfrac{{\rm {\bm \sigma \bf {p}'}}\sigma^{i}}{I\cdot III^{2}\left( {IV+II} \right)} \right)  \\[12pt]
	\times\left\langle {{\rm {\bf {p}'}}\left| {A^{0}} 
		\right|{\rm {\bf {{{p}'''}}}}} \right\rangle \left\langle {{\rm {\bf 
				{{{p}'''}}}}\left| {A^{i}} \right|{\rm {\bf p}}^{IV}} \right\rangle 
	\left\langle {{\rm {\bf p}}^{IV}\left| {A^{0}} \right|{\rm {\bf {{p}''}}}} 
	\right\rangle -\dfrac{I\sigma^{i}\sigma^{k}}{III^{2}\left( {IV+II} 
		\right)} \\ [12pt]
	\times \left\langle {{\rm {\bf {p}'}}\left| {A^{i}} \right|{\rm {\bf 
				{{{p}'''}}}}} \right\rangle \left\langle {{\rm {\bf {{{p}'''}}}}\left| 
		{A^{k}} \right|{\rm {\bf p}}^{IV}} \right\rangle \left\langle {{\rm {\bf 
				p}}^{IV}\left| {A^{0}} \right|{\rm {\bf {{p}''}}}} \right\rangle  \\ [12pt]
	-\dfrac{\sigma^{i}\sigma^{k}II}{IV^{2}\left( {I+III} \right)}\left\langle 
	{{\rm {\bf {p}'}}\left| {A^{0}} \right|{\rm {\bf {{{p}'''}}}}} \right\rangle 
	\left\langle {{\rm {\bf {{{p}'''}}}}\left| {A^{i}} \right|{\rm {\bf 
				p}}^{IV}} \right\rangle \left\langle {{\rm {\bf p}}^{IV}\left| {A^{k}} 
		\right|{\rm {\bf {{p}''}}}} \right\rangle  \\ [12pt]
	+\left( {-\dfrac{{\rm {\bm \sigma \bf {p}'}}\sigma^{i}II}{I\cdot 
			III^{2}IV^{2}}+\dfrac{{\rm {\bm \sigma \bf {{{p}'''}}}}\sigma 
			^{i}II}{III^{2}IV^{2}\left( {I+III} \right)}+\dfrac{{\rm {\bm \sigma 
					\bf p}}^{IV}\sigma^{i}II}{IV^{2}\left( {I+IV} \right)\left( {I+III} 
			\right)\left( {III+IV} \right)}} \right. \\ [12pt]
	+\left. {\dfrac{\sigma^{i}{\rm {\bm \sigma \bf {{p}''}}}}{II\left( {I+IV} 
			\right)\left( {I+III} \right)\left( {III+IV} \right)}} \right)\left\langle 
	{{\rm {\bf {p}'}}\left| {A^{0}} \right|{\rm {\bf {{{p}'''}}}}} \right\rangle 
	\left\langle {{\rm {\bf {{{p}'''}}}}\left| {A^{o}} \right|{\rm {\bf 
				p}}^{IV}} \right\rangle \left\langle {{\rm {\bf p}}^{IV}\left| {A^{i}} 
		\right|{\rm {\bf {{p}''}}}} \right\rangle  \\ [12pt]
	+\left( {-\dfrac{I\sigma^{i}{\rm {\bm \sigma \bf {{p}''}}}}{II\cdot 
			III^{2}IV^{2}}+\dfrac{l\sigma^{i}{\rm {\bm \sigma 
					\bf {{{p}'''}}}}}{III^{2}\left( {III+II} \right)\left( {III+IV} \right)\left( 
			{IV+II} \right)}+\dfrac{I\sigma^{i}{\rm {\bm \sigma \bf p}}^{IV}}{III^{2}\cdot 
			IV^{2}\left( {IV+II} \right)}} \right. \\ [12pt]
	\left. {+\left. {\dfrac{{\rm {\bm \sigma \bf {p}'}}\sigma^{i}}{I\left( {III+II} 
				\right)\left( {III+IV} \right)\left( {IV+II} \right)}} \right)\left\langle 
		{{\rm {\bf {p}'}}\left| {A^{i}} \right|{\rm {\bf {{{p}'''}}}}} \right\rangle 
		\left\langle {{\rm {\bf {{{p}'''}}}}\left| {A^{0}} \right|{\rm {\bf 
					p}}^{IV}} \right\rangle \left\langle {{\rm {\bf p}}^{IV}\left| {A^{0}} 
			\right|{\rm {\bf {{p}''}}}} \right\rangle } \right\} . \\ 
\end{array} \tag{A.2}
\]

\section*{APPENDIX B. Calculation of matrix elements of QED processes with equations for fermions with spinor wave functions}

\begin{enumerate}
	\item {\it { Electron scattering in Coulomb field:}}  $A^{0}\left( x \right)={Ze} \mathord{\left/ {\vphantom {{Ze} {4\pi \left| {{\rm {\bf x}}} \right|}}} \right. \kern-\nulldelimiterspace} {4\pi \left| {{\rm {\bf x}}} \right|}$. The Feynman diagram is presented in Fig.1. Matrix element $S_{fi} $ is

\[ 
	\begin{array}{l}
		\label{appB.1}
S_{fi} =-i\int {d^{4}x} \bar{{F}}_{0}^{+} \left( {x,p_{f} ,s_{f} } 
\right)V_{10} A^{0}F_{0}^{+} \left( {x,p_{i} ,s_{i} } \right) \\ [12pt]
=-\dfrac{i\delta \left( {E_{f} -E_{i} } \right)}{\left( {2\pi } 
	\right)^{2}2E_{i} }\bar{{U}}_{s_{f} } V_{10} \left( {p_{f} ;p_{i} } 
\right)A^{0}\left( q \right)U_{s_{i} }  \\ [12pt]
=i\dfrac{Ze^{2}}{{\rm {\bf q}}^{{\rm {\bf 2}}}}\dfrac{\delta \left( {E_{f} 
		-E_{i} } \right)}{\left( {2\pi } \right)^{2}}\bar{{U}}_{s_{f} } 
\dfrac{1}{2E_{i} }\left( {E_{i} +m+\frac{1}{E_{i} +m}{\rm {\bm \sigma 
			\bf p}}_{{\rm {\bf f}}} {\rm {\bm \sigma \bf p}}_{{\rm {\bf i}}} } \right)U_{s_{i} } ,  \\ 
\end{array} \tag{B.1}
\]

where ${\rm {\bf q}}={\rm {\bf p}}_{{\rm {\bf f}}} -{\rm {\bf p}}_{{\rm {\bf 
		i}}} $, $A^{0}\left( {{\rm {\bf q}}} \right)={4 \pi Ze} \mathord{\left/ {\vphantom 
	{{Ze} {{\rm {\bf q}}^{2}}}} \right. \kern-\nulldelimiterspace} {{\rm {\bf 
		q}}^{2}}$.

In \eqref{appB.1} for the electron we used function $F_{0}^{+} \left( {x,p_{i} 
,s_{i} } \right)$ and the equality of \\ $II=\left( {E_{i} +m} \right)^{1 
\mathord{\left/ {\vphantom {1 2}} \right. \kern-\nulldelimiterspace} 2}$ 
(see (\ref{eq17}), (\ref{eq30}), and (\ref{eq32})).

The expression \eqref{appB.1} coincides with the expression obtained earlier in the 
Foldy-Wouthuysen representation \cite{bib13}--\cite{bib15}. Then, by using 
conventional methods, we can obtain the differential cross-section of Mott 
scattering transforming into Rutherford scattering in non-relativistic case 
(see, for example, Ref. \cite{bib9}).

	\item {\it {Scattering of an electron on a Dirac proton (M{\o}ller scattering):}} The Feynman diagram is presented in Fig. 2.

\[
\begin{array}{l}
	\label{appB.2}
S_{fi} =-i\int {d^{4}xd^{4}y} \bar{{F}}_{0}^{+} \left( {x,p_{f} ,s_{f} } 
\right)V_{1}^{\alpha } F_{0}^{+} \left( {x,p_{i} ,s_{i} } \right)D_{f} 
\left( {x-y} \right)\, \\ [12pt]
\times \bar{{F}}^{+}\left( {y,P_{f} ,S_{f} } \right)\left( {-V_{1} } 
\right)_{\alpha } F^{+}\left( {y,P_{i} ,S_{i} } \right)=-\dfrac{i\delta 
	^{4}\left( {P_{f} -P_{i} +p_{f} -p_{i} } \right)}{\left( {p_{f} -p_{i} } 
	\right)^{2}} \\ [12pt]
\times \dfrac{2\pi }{\sqrt {2p_{i}^{0} 2p_{f}^{0} } \sqrt {2P_{i}^{0} 
		2P_{f}^{0} } }\left( {\bar{{U}}_{s_{f} } V_{1}^{\alpha } \left( {p_{f} 
		;p_{i} } \right)U_{s_{i} } } \right)\,\left( {\bar{{U}}_{S_{f} } V_{1\alpha 
	} \left( {P_{f} ;P_{i} } \right)U_{S_{i} } } \right). \\ 
\end{array} \tag{B.2}
\]

Above, $D_{f} \left( {x-y} \right)$ is a photon propagator. Matrix element 
$S_{fi} $ allows determination of M{\o}ller cross-section.
	\item {\it {Compton scattering of electrons:}} The Feynman diagrams are presented in Fig.3.

We will describe the incident photon with momentum $k^{\mu }$ and 
polarization $\varepsilon^{\mu }$ by plane wave
\[
A^{\mu }\left( {x,k} \right)=\frac{\varepsilon^{\mu }}{\sqrt {2k^{0}\left( 
	{2\pi } \right)^{3}} }e^{-ikx}.
\]
The emitted photon with momentum ${k}'^{\mu }$ and polarization 
${\varepsilon }'^{\mu }$ is described by plane wave
\[
A^{\mu }\left( {y,{k}'} \right)=\frac{\left( {{\varepsilon }'} \right)^{\mu 
}}{\sqrt {2\left( {{k}'} \right)^{0}\left( {2\pi } \right)^{3}} 
}e^{i{k}'y}.
\]

The matrix element of the process is

\[
	\begin{array}{l} 
		\label{appB.3}
S_{fi} =-i\bar{{U}}_{s_{f} } \left\{ {\int 
	{\dfrac{d^{4}zd^{4}yd^{4}{{{p}'''}}}{\left( {2\pi } \right)^{10}\sqrt 
			{2k^{0}2\left( {{k}'} \right)^{0}2p_{i}^{0} 2p_{f}^{0} } } } } \right. \\ [12pt]
\times \left( {e^{ip_{f} y}\,V_{1\mu } \,{\varepsilon }'^{\mu }} 
\right.e^{i{k}'\,y}\,\dfrac{e^{-i{{{p}'''}}y}}{\left( {{{{p}'''}}} 
	\right)^{2}-m^{2}}e^{i{{{p}'''}}z}V_{1\mu } \varepsilon 
^{\nu}e^{-ikz}e^{ip_{i} z} \\ [12pt]
+\left. {e^{ip_{f} y}V_{1\mu } \varepsilon^{\mu 
	}e^{-iky}\dfrac{e^{-i{{{p}''}}y}}{\left( {{{{p}}}} 
		\right)^{2}-m^{2}}e^{i{{{p}'''}}z}V_{1\nu} {\varepsilon 
	}'^{\nu}\,e^{i{k}'\,z}e^{-ip_{i} z}} \right) \\ [12pt]
+\int {d^{4}y\dfrac{1}{\left( {2\pi } \right)^{6}\sqrt {2k^{0}2\left( {{k}'} 
			\right)^{0}2p_{i}^{0} 2p_{f}^{0} } }\left( {e^{ip_{f} y}} \right.V_{2\mu 
		\nu} \,{\varepsilon }'^{\mu }e^{i{k}'\,y}\,\varepsilon 
	^{\nu}e^{-iky}e^{-ip_{i} y}} \\ [12pt]
\left. {+\left. {e^{ip_{f} y}V_{2\mu \nu} \,\varepsilon^{\mu 
		}e^{-iky}{\varepsilon }'^{\nu}e^{i{k}'y}e^{-ip_{i} y}} \right)} \right\} 
U_{s_{i} } = \\ [12pt]
-\dfrac{i\delta^{4}\left( {p_{i} +k-p_{f} -{k}'} \right)}{\left( 
	{2\pi } \right)^{2}\sqrt {2k^{0}2\left( {{k}'} \right)^{0}2p_{i}^{0} 
		2p_{f}^{0} } }\left( {\bar{{U}}_{s_{f} } MU_{s_{i} } } \right), \\ 
\end{array} \tag{B.3}
\]

where
\[
\begin{array}{l}
M=V_{1\mu } \,\left( {p_{f} ;p_{i} +k} \right){\varepsilon }'^{\mu 
}\dfrac{1}{\left( {p_{i} +k} \right)^{2}-m^{2}}V_{1\mu } \,\left( {p_{i} 
	+k;p_{i} } \right)\varepsilon^{\mu } \\ [12pt]
+V_{2\mu \nu} \,\left( {p_{f} ;p_{i} 
	+k;p_{i} } \right){\varepsilon }'^{\mu }\varepsilon^{\nu} \\ [12pt]
+V_{1\mu } \,\left( {p_{f} ;p_{i} -{k}'} \right)\varepsilon^{\mu 
}\dfrac{1}{\left( {p_{i} -{k}'} \right)^{2}-m^{2}}V_{1\mu } \,\left( {p_{i} 
	-{k}';p_{i} } \right){\varepsilon }'^{\mu }\\ [12pt]
+V_{2\mu \nu} \,\left( {p_{f} 
	;p_{i} -{k}';p_{i} } \right)\varepsilon^{\mu }{\varepsilon }'^{\nu}. \\ 
\end{array}
\]

Here, taking into account the conservation of energy-momentum of $\left( 
{p_{i} +k=p_{f} +{k}'} \right)$, operators $V_{1\mu } \,,V_{2\mu \nu } \,$ 
are determined from expressions (\ref{eq35}), \eqref{appA.1}, taken without fields $A^{\mu 
},A^{\mu }A^{\nu}$.

If we select special calibration, in which an initial and final photons are 
transversely polarized in the laboratory system of reference of $\left( {\bf p_{i}}=0, p^{0}_{i}=m \right.$, $\left. \varepsilon ^{0} = \left( \varepsilon' \right) ^{0} =0, \varepsilon {\bf k} = \varepsilon' {\bf k'} =0 \right)$ then the expression for $S_{fi} $ is 
simplified:

\[
	\begin{array}{l}
		\label{appB.4}
M=e^{2}\sqrt {\dfrac{2m+k^{0}-\left( {{k}'} \right)^{0}}{2m}} 2\left\{ 
{{\rm {\bm {\varepsilon }'\bm \varepsilon }}+\dfrac{1}{2k^{0}\left( 
		{2m+k^{0}-k^{0}} \right)}{\rm {\bm \sigma }}\left( {{\rm {\bf k}}-{\rm {\bf 
				{k}'}}} \right){\rm {\bm \sigma {\bm\varepsilon }'\bm\sigma \bf k \bm\sigma \bm\varepsilon 
	}}} \right. \\ [12pt]
+\left. {\dfrac{1}{2\left( {{k}'} \right)^{0}\left( {2m+k^{0}-k^{0\prime}} \right)} {\rm {\bm \sigma }}\left( {{\rm {\bf k}}-{\rm {\bf {k}'}}} 
	\right){\rm {\bm \sigma \bm \varepsilon \bm \sigma \bf {k}' \bm \sigma {\bm \varepsilon }'}}} 
\right\}.  
\end{array} \tag{B.4}
\]

While obtaining the latter expression, we used equalities of \\ $I\equiv \sqrt 
{2m+k^{0}-\left( {{k}'} \right)^{0}} $, 
$II=\sqrt {2m} , III\equiv \sqrt {2m+k^{0}} ,$ or $III\equiv \sqrt {2m-\left( {{k}'} 
\right)^{0}} $, ${\rm {\bm \sigma \bf {p}'}}={\rm {\bm \sigma }}\left( {{\rm 
	{\bf k}}-{\rm {\bf {k}'}}} \right)$, ${\rm {\bm \sigma \bf {{p}''}}}=0$, ${\rm 
{\bm \sigma \bf {{{p}'''}}}}={\rm {\bm \sigma \bf k}}$, or ${\rm {\bm \sigma 
	{{{p}'''}}}}=-{\rm {\bm \sigma \bf {k}'}}$.

Then, by using conventional methods, we can obtain the Klein-Nishina-Tamm 
formula for the differential cross-section of the Compton scattering. 

	\item {\it {Electron-positron pair annihilation:}} The process of electron-positron pair annihilation is corresponded by the diagram in Fig. 3 with substitution of $\varepsilon ,k\to \varepsilon_{1} 
,-k_{1} ,\,\,{\varepsilon }',{k}'\to \varepsilon_{2} ,k_{2} ,\,p_{i} s_{i} 
\to p_{-} s_{-} $, $p_{f} s_{f} \to p_{+} s_{+} $. When recording matrix 
element of $S_{+\,-} $, for the positron, we used the function of $F_{0}^{+} 
\left( {x,p_{+} ,s} \right)$ and $I=\left( {p_{+}^{0} -m} \right)^{1 
\mathord{\left/ {\vphantom {1 2}} \right. \kern-\nulldelimiterspace} 2}$ 
(see Eqs. (\ref{eq17}), (\ref{eq30}), and (\ref{eq32})).

In analogy with Compton scattering, the matrix element of process $S_{+\,-} 
$ is

\[ 
	\begin{array}{l}
		\label{appB.5}
S_{+\,-} =-\dfrac{\delta^{4}\left( {p_{-} +p_{+} -k_{1} -k_{2} } 
\right)}{\left( {2\pi } \right)^{2}\sqrt {2k_{1}^{0} 2k_{2}^{0} \cdot 
	2p_{-}^{0} 2p_{+}^{0} } }\bar{{U}}_{s_{+} } M_{1} U_{s_{-} } ,
\end{array} \tag{B.5}
\]

where operator $M_{1} $, in terms of its structure, considering the above 
substitution, coincides with operator $M$ in the expression of $S_{fi} $ for 
Compton scattering of electrons. The expression of \eqref{appB.5} allows obtaining 
the differential cross-section of the electron-positron pair annihilation, 
which coincides with the cross-section of this process calculated in Dirac 
representation. 
\end{enumerate}

\section*{APPENDIX C. Contributions of Figs. 5(i)--5(l) to the Lamb shift}

\begin{itemize}
	\item {\it {Contribution of Fig.5(i)}}
\[
	\begin{array}{l}
		\label{appC.1}
	e^3 	\left\{ {A_{1} B_{1} C_{1} +B_{1} C_{1} \dfrac{\left( {{\rm {\bf {p}'}}} 
				\right)^{2}-{\rm {\bm \sigma \bf {p}' \bm\sigma k}}}{I\cdot III}} \right.+A_{1} 
		C_{1} \dfrac{{\rm {\bm \sigma \bf {p}' \bm\sigma {{p}''}}}-{\rm {\bm \sigma 
					\bf {p}' \bm\sigma k}}-{\rm {\bm \sigma \bf k \bm\sigma {{p}''}}}+{\rm {\bf 
					k}}^{2}}{III^{2}} \\ [12pt]
		+A_{1} B_{1} \dfrac{\left( {{\rm {\bf {{p}''}}}} \right)^{2}-{\rm {\bm 
					\sigma \bf k \bm\sigma {{p}''}}}}{I\cdot III}+C_{1} \dfrac{\left( {{\rm {\bf {p}'}}} 
			\right)^{2}-2{\rm {\bf {p}'k}}+{\rm {\bf k}}^{2}}{I\cdot III^{3}}\left( 
		{{\rm {\bm \sigma \bf {p}' \bm \sigma {{p}''}}}-{\rm {\bm \sigma \bf {p}' \bm\sigma k}}} 
		\right) \\ [12pt]
		+B_{1} \dfrac{{\rm {\bf k}}^{2}}{I^{2}\cdot III^{2}}{\rm {\bm \sigma 
				\bf {p}' \bm\sigma {{p}''}}} +A_{1} \dfrac{\left( {{\rm {\bf {{p}''}}}} \right)^{2}-2{\rm {\bf {{p}''}k}} +{\rm {\bf k}}^{2}}{I\cdot III^{3}}\left( {{\rm {\bm \sigma 
					\bf {p}' \bm\sigma {{p}''}}}-{\rm {\bm \sigma \bf k \bm\sigma {{p}''}}}} \right) \\ [12pt]
				 +\dfrac{{\rm {\bf k}}^{4}}{I^{2}\cdot III^{2}}{\rm {\bm \sigma \bf {p}'\bm \sigma {{p}''}}}+B_{1} 
		\left( {{\rm {\bm \sigma \bf {p}' \bm\sigma {{p}''}}}-{\rm {\bm \sigma \bf {p}' \bm\sigma 
					k}}} \right) \\  [12pt]
		-B_{1} \dfrac{I^{2}}{III^{2}}\left( {-{\rm {\bm \sigma \bf {p}' \bm\sigma 
					{{p}''}}}+{\rm {\bm \sigma \bf k \bm\sigma {{p}''}}}+{\rm {\bm \sigma \bf {p}' \bm\sigma 
					k}}+3{\rm {\bf k}}^{2}+4{\rm {\bf {p}'{{p}''}}}-4{\rm {\bf {p}'k}}-4{\rm 
				{\bf {{p}''}k}}} \right) \\  [12pt]
		-\dfrac{3III^{2}}{I^{2}}B_{1} {\rm {\bm \sigma \bf {p}' \bm\sigma {{p}''}}}+B_{1} 
		\left( {{\rm {\bm \sigma \bf {p}' \bm\sigma {{p}''}}}-{\rm {\bm \sigma \bf k \bm\sigma 
					{{p}''}}}} \right) \\ [12pt]
			+\dfrac{\left( {{\rm {\bf {{p}''}}}} \right)^{2}-2{\rm {\bf 
					{{p}''}k}}+{\rm {\bf k}}^{2}}{III^{2}}\left( {\left( {{\rm {\bf {p}'}}} 
			\right)^{2}-{\rm {\bm \sigma \bf {p}' \bm\sigma k}}} \right) \\  [12pt]
		-\dfrac{3I^{2}}{III^{4}}\left( {\left( {{\rm {\bf {p}'}}} \right)^{2}-2{\rm 
				{\bf {p}'k}}+{\rm {\bf k}}^{2}} \right)\left( {\left( {{\rm {\bf {{p}''}}}} 
			\right)^{2}-2{\rm {\bf {{p}''}k}}+{\rm {\bf k}}^{2}} 
		\right)-\dfrac{3k^{2}}{I^{2}}{\rm {\bm \sigma \bf {p}' \bm\sigma {{p}''}}} \\  [12pt]
		+\left. {\dfrac{\left( {{\rm {\bf {p}'}}} \right)^{2}-2{\rm {\bf 
						{p}'k}}+{\rm {\bf k}}^{2}}{III^{2}}\left( {\left( {{\rm {\bf {{p}''}}}} 
				\right)^{2}-{\rm {\bm \sigma \bf k \bm\sigma {{p}''}}}} \right)} \right\} \\  [12pt]
		\times \dfrac{A^0 \left( \bf q \right)}{\left( {\left( {k^{0}} \right)^{2}-{\rm {\bf 
						k}}^{2}-2\left( {{\rm {\bf {p}'}}} \right)^{0}k^{0}+2{\rm {\bf {p}'k}}} 
			\right)\left( {\left( {k^{0}} \right)^{2}-{\rm {\bf k}}^{2}-2\left( {{\rm 
						{\bf {{p}''}}}} \right)^{0}k^{0}+2{\rm {\bf {{p}''}k}}} \right)}, \\ [12pt]
		A_{1} =I^{2}+III^{2}-2m-\dfrac{\left( {{\rm {\bf {p}'}}} 
			\right)^{2}}{I\left( {I+III} \right)}-\dfrac{\left( {{\rm {\bf {p}'}}} 
			\right)^{2}-2{\rm {\bf {p}'k}}+{\rm {\bf k}}^{2}}{III\left( {I+III} 
			\right)}, \\ [12pt]
		C_{1} =I^{2}+III^{2}-2m-\dfrac{\left( {{\rm {\bf {{p}''}}}} 
			\right)^{2}}{I\left( {I+III} \right)}-\dfrac{\left( {{\rm {\bf {{p}''}}}} 
			\right)^{2}-2{\rm {\bf {{p}''}k}}+{\rm {\bf k}}^{2}}{III\left( {I+III} 
			\right)}, \\ [12pt]
		B_{1} =2III^{2}-2m-\dfrac{\left( {{\rm {\bf {p}'}}} \right)^{2}-2{\rm {\bf 
					{p}'k}}+{\rm {\bf k}}^{2}}{2III^{2}}-\dfrac{\left( {{\rm {\bf {{p}''}}}} 
			\right)^{2}-2{\rm {\bf {{p}''}k}}+{\rm {\bf k}}^{2}}{2III^{2}}. \tag{C.1}
	\end{array}
	\]
		\item {\it {Contribution of Fig. 5(j)}}
	\[ 
	\begin{array}{l}
			e^3 	\left\{ {DE+D\dfrac{{\rm {\bm \sigma \bf {p}'\bm \sigma {{p}''}}}-{\rm {\bm \sigma \bf
							k \bm \sigma {{p}''}}}}{I\cdot III^{3}}} \right.-D\dfrac{{\rm {\bm \sigma 
						\bf {p}' \bm\sigma {{p}''}}}-{\rm {\bm \sigma \bf k \bm\sigma {{p}''}}}-{\rm {\bm \sigma 
						\bf {p}'\bm\sigma k}}+{\rm {\bf k}}^{2}}{III^{3}\left( {I+III} 
				\right)} \\ [12pt]
			- D\dfrac{\left( {{\rm {\bf {{p}''}}}} \right)^{2}-{\rm {\bm \sigma 
						\bf k \bm\sigma {{p}''}}}}{2I\cdot III^{3}}+ E\dfrac{\left( {{\rm {\bf {p}'}}} \right)^{2}-{\rm {\bm \sigma \bf {p}'\bm \sigma 
						k}}}{I\cdot III}+\dfrac{{\rm {\bf k}}^{2}}{2I^{2}\cdot III^{4}}{\rm {\bm 
					\sigma \bf {p}' \bm\sigma {{p}''}}} \\ [12pt]
				- \dfrac{\left( {{\rm {\bf {p}'}}} 
				\right)^{2}-2{\rm {\bf {p}'k}}+{\rm {\bf k}}^{2}}{I\cdot III^{4}\left( 
				{I+III} \right)}\left( {{\rm {\bm \sigma \bf {p}' \bm\sigma {{p}''}}}-{\rm {\bm 
						\sigma \bf {p}' \bm\sigma k}}} \right) \\  [12pt]
\end{array}
\]
\[
\begin{array}{l}
	\label{appC.2}	
			+\dfrac{\left( {{\rm {\bf {p}'}}} \right)^{2}-{\rm {\bm \sigma \bf {p}' \bm\sigma 
						k}}}{III^{2}}-\dfrac{3I^{2}}{III^{4}}\left( {\left( {{\rm {\bf {p}'}}} 
				\right)^{2}-2{\rm {\bf {p}'k}}+{\rm {\bf k}}^{2}} 
			\right)+\dfrac{3}{2I^{2}}{\rm {\bm \sigma \bf {p}' \bm\sigma {{p}''}}} \\ [12pt]
			-\dfrac{{\rm {\bm \sigma \bf {p}' \bm\sigma {{p}''}}}-{\rm {\bm \sigma \bf k \bm\sigma 	{{p}''}}}}{2III^{2}}- \dfrac{{\rm {\bm \sigma \bf {p}' \bm\sigma {{p}''}}}-{\rm {\bm \sigma \bf {p}'\bm \sigma 
						k}}}{2III^{2}} \\ [12pt]
					+ \left. {\dfrac{I^{2}}{2III^{4}}\left( {-{\rm {\bm \sigma 
							\bf {p}' \bm\sigma {{p}''}}}+{\rm {\bm \sigma \bf k \bm\sigma {{p}''}}}+{\rm {\bm \sigma 
							\bf {p}' \bm\sigma k}}+3{\rm {\bf k}}^{2}+4{\rm {\bf {p}'{{p}''}}}-4{\rm {\bf 
							{p}'k}}-4{\rm {\bf {{p}''}k}}} \right)} \right\}  \\  [12pt]
			\times \dfrac{A^0 \left( \bf q \right)}{\left( {k^{0}} \right)^{2}-{\rm {\bf k}}^{2}-2\left( {{\rm 
						{\bf {p}'}}} \right)^{0}k^{0}+2{\rm {\bf {p}'k}}}. \\ [12pt]
				D=III^{2}+I^{2}-2m-\dfrac{\left( {{\rm {\bf {p}'}}} \right)^{2}}{I\cdot 
				III}+\dfrac{2{\rm {\bf {p}'k}}}{III\left( {I+III} \right)}-\dfrac{{\rm {\bf 
						k}}^{2}}{III\left( {I+III} \right)}, \\ [12pt]
			E=-1-\dfrac{{\rm {\bf k}}^{2}}{2III^{2}\left( {I+III} 
				\right)^{2}}+\dfrac{{\rm {\bf k}}^{2}}{2III^{3}\left( {I+III} 
				\right)}+\dfrac{{\rm {\bf {p}'k}}}{III^{2}\left( {I+III} 
				\right)^{2}}  \\ [12pt]
			-\dfrac{{\rm {\bf {{p}''}k}}}{III^{3}\left( {I+III} \right)} -\dfrac{\left( {{\rm {\bf {p}'}}} \right)^{2}}{2III^{2}\left( {I+III} 
				\right)^{2}}  \\ [12pt]
			+\left( {{\rm {\bf {{p}''}}}} \right)^{2}\left( 
			{\dfrac{1}{2III^{3}\left( {I+III} \right)}-\dfrac{1}{2I\cdot III\left( {I+III} 
					\right)^{2}}} \right). \tag{C.2} 
		\end{array}
		\]

	\item {\it {Contribution of Fig. 5(k)}}
\[ 
	\begin{array}{l}
		\label{appC.3}
			e^3 \left\lbrace  {D_{1} E_{1} +D_{1} \dfrac{{\rm {\bm \sigma \bf {p}'\bm \sigma 
							{{p}''}}}-{\rm {\bm \sigma \bf {p}' \bm \sigma k}}}{I\cdot III^{3}}} \right.-D_{1} 
			\dfrac{\left( {{\rm {\bf {p}'}}} \right)^{2}-{\rm {\bm \sigma \bf {p}' \bm \sigma 
						k}}}{2I\cdot III^{3}} \\ [12pt]
					-D_{1} \dfrac{\left( {{\rm {\bm \sigma \bf {p}'\bm \sigma 
							{{p}''}}}-{\rm {\bm \sigma \bf {p}' \bm \sigma k}}-{\rm {\bm \sigma \bf k \bm \sigma 
							{{p}''}}}+{\rm {\bf k}}^{2}} \right)}{III^{3}\left( {I+III} \right)} 
			+E_{1} \dfrac{\left( {{\rm {\bf {{p}''}}}} \right)^{2}-{\rm {\bm \sigma 
						\bf k \bm \sigma {{p}''}}}}{III\cdot I}  \\ [12pt]
					+\dfrac{{\rm {\bf k}}^{2}}{2I^{2}\cdot 
				III^{4}}{\rm {\bm \sigma \bf {p}'\bm \sigma {{p}''}}}-\dfrac{\left( {{\rm {\bf 
							{{p}''}}}} \right)^{2}-2{\rm {\bf {{p}''}k}}+{\rm {\bf k}}^{2}}{I\cdot 
				III^{4}\left( {I+III} \right)}\left( {{\rm {\bm \sigma \bf {p}'\bm \sigma 
						{{p}''}}}-{\rm {\bm \sigma \bf k \bm \sigma {{p}''}}}} \right) \\ [12pt]
			-\dfrac{{\rm {\bm \sigma \bf {p}'\bm \sigma {{p}''}}}-{\rm {\bm \sigma \bf {p}' \bm \sigma 
						k}}}{2III^{2}}-\dfrac{3I^{2}}{III^{4}}\left( {\left( {{\rm {\bf {{p}''}}}} 
				\right)^{2}-2{\rm {\bf {{p}''}k}}+{\rm {\bf k}}^{2}} 
			\right)  \\ [12pt]
			+\dfrac{3}{2I^{2}}{\rm {\bm \sigma \bf {p}'\bm \sigma {{p}''}}}+\dfrac{\left( 
				{{\rm {\bf {{p}''}}}} \right)^{2}-{\rm {\bm \sigma \bf k \bm \sigma 
						{{p}''}}}}{III^{2}} -\dfrac{{\rm {\bm \sigma \bf {p}'\bm \sigma {{p}''}}}-{\rm {\bm \sigma \bf k \bm \sigma 	{{p}''}}}}{2III^{2}}  \\ [12pt]
					+\left. {\dfrac{I^{2}}{2III^{4}}\left( {-{\rm {\bm \sigma 
						\bf {p}' \bm \sigma {{p}''}}}+{\rm {\bm \sigma \bf k \bm \sigma {{p}''}}}+{\rm {\bm \sigma \bf
							{p}' \bm\sigma k}}+3{\rm {\bf k}}^{2}+4{\rm {\bf {p}'{{p}''}}}-4{\rm {\bf 
							{p}'k}}-4{\rm {\bf {{p}'}'k}}} \right)}  \right\rbrace \\ [12pt]
			\times \dfrac{A^0 \left( \bf q \right) }{\left( {k^{0}} \right)^{2}-{\rm {\bf k}}^{2}-2\left( {{\rm 
						{\bf {{p}''}}}} \right)^{0}k^{0}+2{\rm {\bf {{p}''}k}}}, \\ 
				D_{1} =I^{2}+III^{2}-2m-\dfrac{\left( {{\rm {\bf {{p}''}}}} 
				\right)^{2}}{I\cdot III}+\dfrac{2{\rm {\bf {{p}''}k}}}{III\left( {I+III} 
				\right)}-\dfrac{{\rm {\bf k}}^{2}}{III\left( {I+III} \right)}, \\ [12pt]
			E_{1} =-1+\dfrac{{\rm {\bf k}}^{2}}{2III^{3}\left( {I+III} 
				\right)}-\dfrac{{\rm {\bf k}}^{2}}{2III^{2}\left( {I+III} 
				\right)^{2}}+\dfrac{{\rm {\bf {{p}''}k}}}{III^{2}\left( {I+III} 
				\right)^{2}} \\ [12pt]
			-\dfrac{{\rm {\bf {p}'k}}}{III^{3}\left( {I+III} \right)} -\dfrac{\left( {{\rm {\bf {{p}''}}}} \right)^{2}}{2III^{2}\left( {I+III} 
				\right)^{2}} \\ [12pt]
			+\left( {{\rm {\bf {p}'}}} \right)^{2}\left( 
			{\dfrac{1}{2III^{3}\left( {I+III} \right)}-\dfrac{1}{2III\cdot I\left( {I+III} 
					\right)^{2}}} \right). \tag{C.3}
		\end{array}
		\]
\item {\it {Contribution of Fig. 5(l)}}
\[
	\begin{array}{l}
		\label{appC.4}
	e^3 \left\lbrace \left( {{\rm {\bf {p}'}}} \right)^{2}\left[ {\dfrac{1}{2III^{3}\left( 
			{I+III} \right)^{3}}-\dfrac{1}{2I\cdot III^{3}\left( {I+III} 
			\right)^{2}}-\dfrac{1}{2I^{2}\cdot III\left( {I+III} \right)^{3}}} \right]\right.  
	\\ [12pt]
	+\left( {{\rm {\bf {{p}''}}}} \right)^{2}\left[ {\dfrac{1}{2III^{3}\left( 
			{I+III} \right)^{3}}-\dfrac{1}{2I\cdot III^{3}\left( {I+III} 
			\right)^{2}}-\dfrac{1}{2I^{2}\cdot III\left( {I+III} \right)^{3}}} \right] 
\end{array}
\]
\[	
\begin{array}{l}
		+{\rm {\bm \sigma \bf {p}'\bm\sigma {{p}''}}}\left[ {-\dfrac{2}{I\cdot 
			III^{4}\left( {I+III} \right)}+\dfrac{1}{III^{4}\left( {I+III} 
			\right)^{2}}+\dfrac{1}{I^{2}\cdot III^{4}}} \right] \\ [12pt]
	+{\rm {\bm \sigma \bf k\bm\sigma {{p}''}}}\left[ {\dfrac{1}{2I\cdot III^{3}\left( 
			{I+III} \right)^{2}}-\dfrac{1}{III^{4}\left( {I+III} 
			\right)^{2}}+\dfrac{1}{I\cdot III^{4}\left( {I+III} \right)}} \right] \\ [12pt]
	+{\rm {\bm \sigma \bf {p}'\bm\sigma k}}\left[ {\dfrac{1}{2I\cdot III^{3}\left( 
			{I+III} \right)^{2}}-\dfrac{1}{III^{4}\left( {I+III} 
			\right)^{2}}+\dfrac{1}{I\cdot III^{4}\left( {I+III} \right)}} \right] \\ [12pt]
		- {\rm 	{\bf {p}'k}}\dfrac{1}{III^{3}\left( {I+III} \right)^{3}} -{\rm {\bf {{p}''}k}}\dfrac{1}{III^{3}\left( {I+III} \right)^{3}}+\dfrac{{\rm 
			{\bf k}}^{2}}{III^{3}\left( {I+III} \right)^{3}}\\ [12pt]
		\left. + \dfrac{{\rm {\bf k}}^{2}}{III^{4}\left( {I+III} \right)^{2}}-\dfrac{3I^{2}}{III^{4}}\right\rbrace A^0 \left( \bf q \right) . \\ 
\end{array} \tag{C.4}
\]	
\end{itemize}


\begin{thebibliography}{99}
\bibitem{bib1} P. A. M. Dirac, {\it The Principles of Quantum Mechanics} (Oxford University Press, 1930).

\bibitem{bib2} V. P. Neznamov, {\it Theoret. and Math. Phys.} { \bf{197}}:3, 1823 (2018).

\bibitem{bib3} V. P. Neznamov and I. I. Safronov, {\it  Exp. Theor. Phys.}  {\bf{127}}:647 (2018), arXiv: 1809.08940 [gr-qc].

\bibitem{bib4} V. P. Neznamov, I. I. Safronov and V. E. Shemarulin, {\it  Exp. Theor. Phys.} {\bf{127}}:684 (2018), arXiv: 1810.01960 [gr-qc].

\bibitem{bib5} V. P. Neznamov, I.I. Safronov and V. E. Shemarulin, {\it J. Exp. Theor. Phys.}  {\bf{128}}: 64 (2019), arXiv: 1904.05791 [gr-qc].

\bibitem{bib6} V. P. Neznamov and I. I. Safronov, {\it J. Exp. Theor. Phys.}  {\bf{128}}: 672 (2019), arXiv: 1907.03579 [physics.gen-ph].

\bibitem{bib7} G. Gabrielse, D. Hanneke, T. Kinoshita, M. Noi and B. Odom, {\it Phys. Rev. Lett.} {\bf{97}}, 030809 (2006).

\bibitem{bib8} W. Heitler, {\it The quantum theory of radiation } (At the Clarendon Press, Oxford, 1954).

\bibitem{bib9} J. Bjorken and S. Drell, {\it Relativistic quantum mechanics} (McGraw-Hill Book Company, 1964).

\bibitem{bib10} Ya. B. Zeldovich and V. S. Popov, {\it Sov. Phys. Usp.} {\bf{14}}, 673-694 (1972).

\bibitem{bib11} P. A. I. Dirac, {\it Lectures on quantum field theory}, Belfer Graduate School of Science (Yeshiva University, New York, 1967).

\bibitem{bib12} A. I. Akhiyezer and V. B. Berestetsky, {\it Quantum electrodynamics} (Nauka, Moscow, 1969).

\bibitem{bib13} V. P. Neznamov, {\it VANT, Ser. ''Theoretical and applied physics''} {\bf{1}}, 3 (1989).

\bibitem{bib14} V. P. Neznamov, {\it VANT, Ser. ''Theoretical and applied physics''}, {\bf{1-2}}, 41 (2004).

\bibitem{bib15} V. P. Neznamov, {\it Phys. Part. Nucl.}  {\bf{37}}, 152 (2006), arXiv: hep-th/0411050.

\bibitem{bib16} E.I. Lifshitz and L. P. Pitaevskii, {\it Relativistic quantum theory}, part 2 (Nauka, Moscow, 1971).

\bibitem{bib17} L. S.Hostler, {\it J. Math. Phys.} {\bf{26}}, 1348 (1985).

\bibitem{bib18} G-J. Ni, {\it Rel. Grav. Cosmol.} {\bf{1}} 123-136 (2004), arXiv: 0308038v1.

\bibitem{bib19} G-J. Ni, S. Chen, S. Lou and J. Xu, arXiv: 1007.3051v1.

\bibitem{bib20} N. Debergh, J-P. Petit and G. D'Agostini, arXiv: 1809.05046v2.
\end{thebibliography}
\end{document}